\documentclass[12pt,twoside,letterpaper]{article}
\pdfoutput=1
\usepackage[margin=1.0in]{geometry}
\usepackage{wrapfig,rotating,amssymb,amsmath,here,dsfont}
\usepackage{multirow,float,cite,color,eso-pic}
\usepackage{ifpdf}
\ifpdf
\usepackage{url,boxedminipage}
\usepackage{graphicx}
\definecolor{rltred}{rgb}{0.75,0,0}
\definecolor{rltgreen}{rgb}{0,0.5,0}
\definecolor{rltblue}{rgb}{0,0,0.75}
\usepackage[pagebackref,colorlinks,urlcolor=rltblue,filecolor=rltgreen%
,linkcolor=rltred,pdftitle={Testing hypotheses in particle physics: plots of p0 versus p1}%
,pdfsubject={pdfsubject},pdfkeywords={pdfkeywords}%
,pdfauthor={Luc Demortier <luc.demortier@rockefeller.edu>%
, Louis Lyons <l.lyons1@physics.ox.ac.uk>}%
,pdfpagemode={UseOutlines},pdfpagelayout=SinglePage%
,pdfstartview={Fit}]{hyperref}
\DeclareGraphicsExtensions{.pdf,.jpg,.jpeg}
\else
\usepackage{graphicx}
\DeclareGraphicsExtensions{.eps,.ps,.eps.gz,.ps.gz}
\newcommand{\href}[2]{#2}
\usepackage[dvips,pagebackref]{hyperref}
\fi

\newenvironment{Displaylist}[1]%
{\begin{list}{}%
{\settowidth{\labelwidth}{\rm #1}%
\setlength{\leftmargin}{\labelwidth}%
\addtolength{\leftmargin}{\labelsep}%
}}%
{\end{list}}

\begin{document}
%%%%%%%%%%%%%%%%%%%%%%%%%%%%%%%%%%%%%%%%%%%%%%%%%%%%%%%%%%%%%%%%%%%%%%%%%%%%%%%%
\thispagestyle{empty}
\begin{Large}
\begin{center}
{\bf Testing Hypotheses in Particle Physics: }\\
{\bf Plots of $\mathbf{p_{0}}$ Versus $\mathbf{p_{1}}$}
\end{center}
\end{Large}
\vspace*{2mm}

\begin{center}
\href{mailto:luc.demortier@cern.ch,l.lyons1@physics.ox.ac.uk}
{Luc Demortier$^{\,\dagger}$, Louis Lyons$^{\,\ddagger}$}\\[2mm]
$^{\dagger}${\it Laboratory of Experimental High Energy Physics}\\
{\it The Rockefeller University, New York, NY 10065, USA}\\[2mm]
$^{\ddagger}${\it Blackett Laboratory}\\
{\it Imperial College, London SW7 2BW, UK}\\[5mm]
{\rm \today}
\end{center}
\vspace*{5mm}
\begin{abstract}
For situations where we are trying to decide which of two hypotheses $H_{0}$ and 
$H_{1}$ provides a better description of some data, we discuss the usefulness of 
plots of $p_{0}$ versus $p_{1}$, where $p_{i}$ is the $p$-value for testing 
$H_{i}$.  They provide an interesting way of understanding the difference between 
the standard way of excluding $H_{1}$ and the $CL_{s}$ approach; the Punzi 
definition of sensitivity; the relationship between $p$-values and likelihood 
ratios; and the probability of observing misleading evidence.  They also help
illustrate the Law of the Iterated Logarithm and the Jeffreys-Lindley paradox.
\end{abstract}

%%%%%%%%%%%%%%%%%%%%%%%%%%%%%%%%%%%%%%%%%%%%%%%%%%%%%%%%%%%%%%%%%%%%%%%%%%%%%%%%
\section{Introduction}
%%%%%%%%%%%%%%%%%%%%%%%%%%%%%%%%%%%%%%%%%%%%%%%%%%%%%%%%%%%%%%%%%%%%%%%%%%%%%%%%
Very often in particle physics we try to see whether some data are consistent 
with the standard model (SM) with the currently known particles (call this 
hypothesis $H_{0}$), or whether it favors a more or less specific form of new 
physics in addition to the SM background ($H_{1}$). This could be, for example, 
a particular form of leptoquark with a well-defined mass; or with a mass in some 
range (e.g. 50 to 1000 GeV).  In the first case there are no free parameters and
$H_{1}$ is described as being `simple', while in the latter case, because of the 
unspecified leptoquark mass, $H_{1}$ is `composite'.

If the only free parameter in the alternative hypothesis $H_{1}$ is the mass of 
some new particle, we can test each mass in $H_{1}$ separately against $H_{0}$, 
in which case we are comparing two simple hypotheses.  However, the ensemble of 
different possible masses in the overall procedure (known as a `raster scan'~\cite{Lyons2014}) 
makes $H_{1}$ composite.  Insight into this type of situation is facilitated by
two-dimensional `$p$-value plots', where the significance of possible 
observations under the null hypothesis is plotted against their significance 
under various values of the free parameter in the alternative 
hypothesis~\cite{Thompson2007}.  The purpose of this article is to use such 
plots to explore various aspects of hypothesis testing in particle 
physics\footnote{In this article we concentrate on hypothesis testing 
procedures pertaining to discovery claims in search experiments (not necessarily
in particle physics).  We do not consider other uses of hypothesis testing, such 
as in particle physics event selection for instance.  The desiderata are 
slightly different.}.

We begin in section~\ref{sec:TestingTypes} by recapitulating the types of 
hypothesis testing familiar from the statistics literature, and contrasting these 
with the practice in particle physics.  Section~\ref{sec:Pvalues} introduces 
$p$-value plots and uses them to discuss the $CL_{s}$ criterion, upper limits,
fixed-hypothesis contours, and the Punzi definition of sensitivity.  The 
probabilities for observations to fall into various regions of a $p$-value plot
are derived in section~\ref{sec:ErrorRates}, together with the error rates and 
power of a particle physics test.  Likelihood ratios form the subject of
section~\ref{sec:LikelihoodRatios}, where they are compared to $p$-values and 
used to plot contours and to compute probabilities of misleading evidence.  
Two famous $p$-value puzzles are described in section~\ref{sec:FamousPuzzles}.
Section~\ref{sec:NuisanceParameters} contains remarks on the effect of 
nuisance parameters, and our conclusions and recommendations appear in 
section~\ref{sec:Conclusion}.  An appendix provides technical details about 
the relationship between $CL_{s}$ and Bayesian upper limits.

%%%%%%%%%%%%%%%%%%%%%%%%%%%%%%%%%%%%%%%%%%%%%%%%%%%%%%%%%%%%%%%%%%%%%%%%%%%%%%%%
\section{Types and outcomes of hypothesis testing}
\label{sec:TestingTypes}
%%%%%%%%%%%%%%%%%%%%%%%%%%%%%%%%%%%%%%%%%%%%%%%%%%%%%%%%%%%%%%%%%%%%%%%%%%%%%%%%
When using observed data to test one or more hypotheses, the first step is to 
design a test statistic $T$ that summarizes the relevant properties of the data.  
The observed value $t$ of $T$ is then referred to its probability distribution 
under each specified hypothesis in order to assess evidence.  The form of the 
test statistic depends on the type of test one is interested in.

Comparisons of data with {\em a single hypothesis} are performed via `goodness 
of fit' tests.  An example of this is the $\chi^{2}$ test, which generally 
requires the data to be binned, and where $T$ is equal to the sum of the squares 
of the numbers of standard deviations between observed and expected bin contents.  
Another well-known technique, which does not require binning, is the 
Kolmogorov-Smirnov test, where $T$ is constructed from the expected and observed 
cumulative distributions of the data.  There are many other 
techniques~\cite{DAgostino1986,Williams2010}.  The outcome of a goodness-of-fit
test is either `Reject' or `Fail to reject' the hypothesis of interest.

Comparison of the data with more than one hypothesis in order to decide which is
favored is known as `hypothesis testing'.  If there are {\em just two simple 
hypotheses} $H_{0}$ and $H_{1}$, the appropriate 
framework is Neyman-Pearson hypothesis testing.  The optimal test statistic $T$
in this case is the likelihood ratio for the two hypotheses, or a one-to-one 
function of it~\footnote{In the case of a counting experiment, the number of 
observed counts $n$ is typically a one-to-one function of the likelihood ratio
for the `signal+background' and `background-only' hypotheses ($H_{1}$ and $H_{0}$
respectively).}.  The outcome of a Neyman-Pearson test is either `Reject $H_{0}$ 
and accept $H_{1}$,' or `Accept $H_{0}$ and reject $H_{1}$.'

In particle physics it often happens that we need to consider additional possible
outcomes of a test.  In the leptoquark example, an observed signal could be due
to something entirely different from a leptoquark: some new physics that we did
not anticipate, or a systematic bias that we did not model.  Hence we may need 
to reject both $H_{0}$ and $H_{1}$ in favor of a third, unspecified hypothesis.  
On the other hand it may also happen that the data sample does not allow us to 
reject either $H_{0}$ or $H_{1}$~\cite{Lehmann1957}.  This leads to the formulation 
of a `double test' of two hypotheses $H_{0}$ and $H_{1}$, which are 
independently tested, resulting in four possible outcomes:
\begin{enumerate}
\item Fail to reject $H_{0}$, and reject $H_{1}$.  This is referred to as 
      `$H_{1}$ excluded,' and in a frequentist approach the rejection of $H_{1}$
      is valid at some level of confidence, typically 95\%.
\item Fail to reject $H_{0}$ and fail to reject $H_{1}$ (`No decision').
\item Reject $H_{0}$, and fail to reject $H_{1}$.  This corresponds to `Discovery
      of $H_{1}$.' In a frequentist approach the rejection of $H_{0}$ is valid 
      at some confidence level, which in particle physics is usually much higher 
      than the confidence level used for excluding $H_{1}$.  Typically the 
      significance level, defined as one minus the confidence level, is set at 
      $2.87\times 10^{-7}$ for rejecting $H_{0}$.  This is the area under a 
      Gaussian tail, starting five standard deviations away from the mean.
\item Reject both $H_0$ and $H_1$.
\end{enumerate}
Often a likelihood ratio is used as the test statistic $T$ for a double test.

For given $H_0$ and for fixed values of the parameters in $H_1$, we can plot 
the probability density functions (pdf's) of $T$, assuming (a) that hypothesis 
$H_{0}$ is true, or (b) that $H_{1}$ is true.  Three possible situations are 
shown in figure~\ref{fig:pdfs}.  In (a), the two hypotheses are hard to 
distinguish as the pdf's lie almost on top of each other; this could happen if 
$H_{1}$ involved a new particle that was only very weakly produced.  In (b), 
the pdf's still overlap to some extent, but distinguishing between the two 
hypotheses may be possible for some data sets.  Finally (c) shows a situation 
where it is relatively easy to choose between the hypotheses.

%%%%%%%%%%%%%%%%%%%%%%%%%%%%%%%%%%%%%%%%%%%%%%%%%%%%%%%%%%%%%%%%%%%%%%%%%%%%%%%%
\section{\texorpdfstring{$\mathbf{p}$-Values}{p-Values}}
\label{sec:Pvalues}
\begin{figure}[H]
\begin{center}
\includegraphics[width=\textwidth]{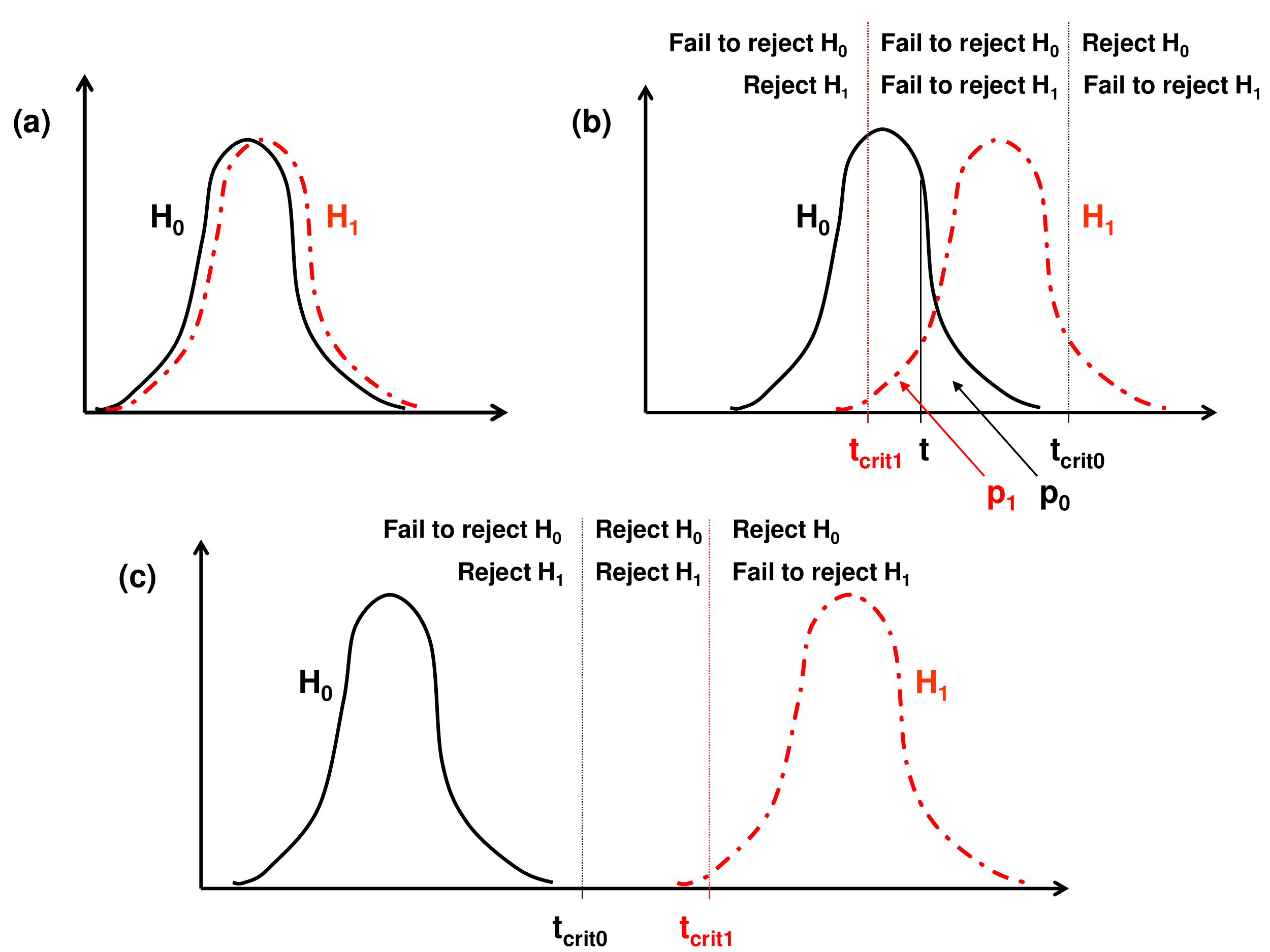}
\caption{Probability density functions (pdf's) of a test statistic $T$ under two 
hypotheses $H_{0}$ (solid) and $H_{1}$ (dot-dashed).  In (a) the separation of 
the pdf's is small, making the two hypotheses almost indistinguishable.  The 
separation in (b) is such that the data can exclude $H_{1}$, reject neither
$H_{0}$ nor $H_{1}$, or reject $H_{0}$.  Given an observed value $t$ of 
the test statistic $T$, diagram (b) illustrates the definitions of $p_{0}$ 
and $p_{1}$: $p_{0}$ is the tail area under $H_{0}$ in the direction of $H_{1}$, 
whereas $p_{1}$ is the tail area under $H_{1}$ in the direction of $H_{0}$. 
The quantities $t_{\rm crit0}$ and $t_{\rm crit1}$ are the critical values of 
$T$, beyond which the data are considered incompatible with $H_0$ and $H_{1}$
respectively; the value of $p_0$ at $T=t_{\rm crit0}$ is $\alpha_{0}$.  The 
$p_{1}$ value at this value of $T$ is denoted by $\beta_{0}$.  In (c) the pdf's 
are far apart and the observed value of $T$ will always reject at least one 
hypothesis.
\label{fig:pdfs}}
\end{center}
\end{figure}

%%%%%%%%%%%%%%%%%%%%%%%%%%%%%%%%%%%%%%%%%%%%%%%%%%%%%%%%%%%%%%%%%%%%%%%%%%%%%%%%
The degree to which the data are unexpected for a given hypothesis can be 
quantified via the $p$-value. This is the fractional area in the tail of the 
relevant pdf, with a value of $t$ at least as extreme as that in the data.  In
tests involving two hypotheses, it is conventional to use the one-sided tail in 
the direction of the other hypothesis.  For the examples shown in 
figure~\ref{fig:pdfs}, this corresponds to $p_{0}$ being the right-hand tail of 
$H_{0}$ and $p_{1}$ the left-hand tail of $H_{1}$.~\footnote{In this paper we do 
not consider problems in which it is desired to reject $H_{0}$ when the data 
statistic $t$ falls in the extreme left-tail of the $H_{0}$ pdf (see 
figure~\protect\ref{fig:pdfs}), or to reject $H_{1}$ when $t$ is very large.  
Our $p$-values are one-sided and would therefore be close to unity in these 
cases.}  In the extreme case where $H_{0}$ and $H_{1}$ coincide (and where $t$ 
is continuous rather than discrete), $p_{0} + p_{1} =1$.

%%%%%%%%%%%%%%%%%%%%%%%%%%%%%%%%%%%%%%%%%%%%%%%%%%%%%%%%%%%%%%%%%%%%%%%%%%%%%%%%
\subsection{Regions in the \texorpdfstring{$\mathbf{(p_0,p_1)}$}{(p0,p1)} plane}
\begin{figure}[p]
\begin{center}
\includegraphics[width=\textwidth]{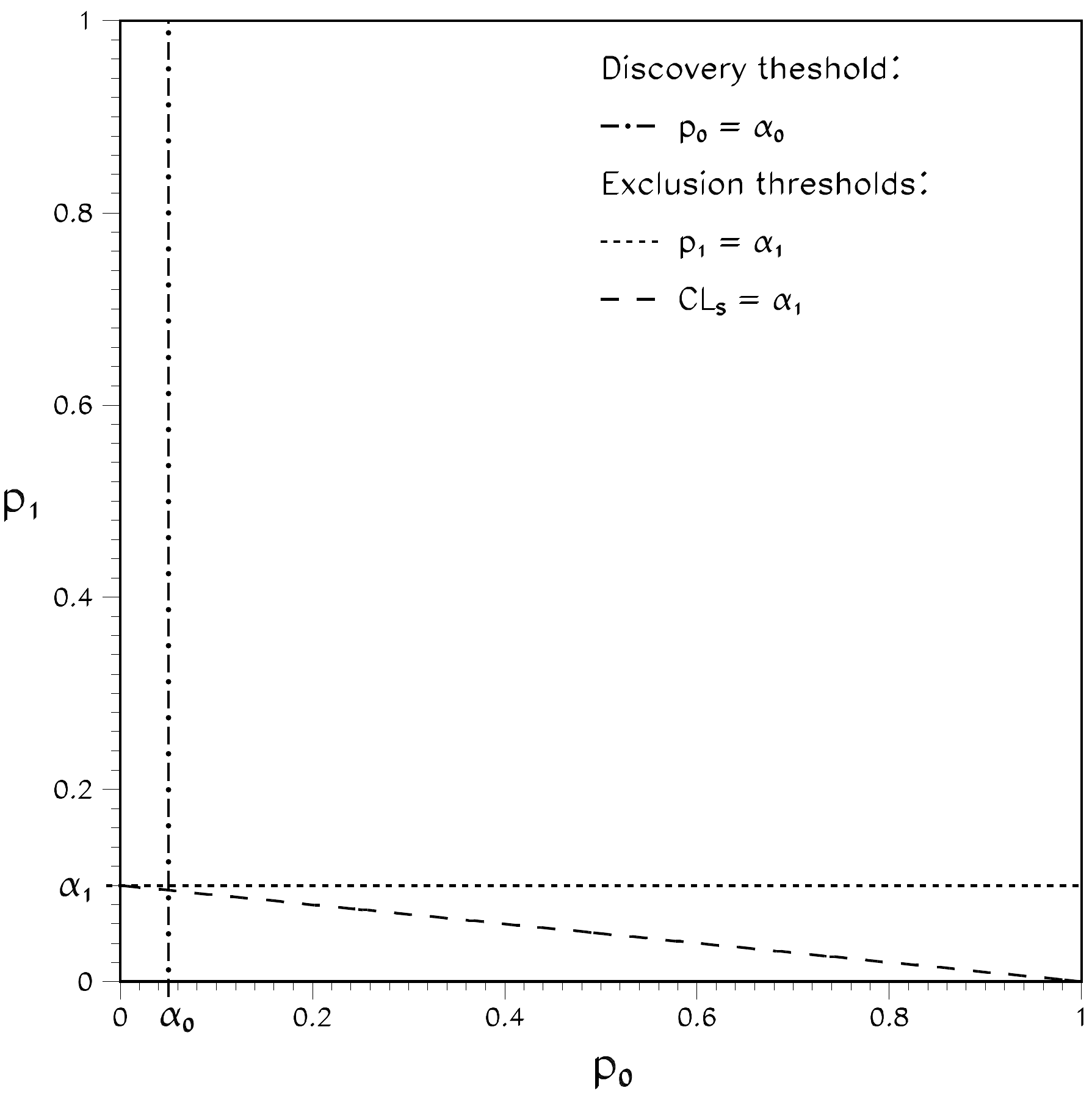}
\caption{Plot of $p_{0}$ versus $p_{1}$ with three lines: The vertical line at 
$p_{0}=\alpha_{0}$ is an example of a cut used for rejecting $H_{0}$ when $p_{0}$ 
is small.  Correspondingly, the horizontal line at $p_{1}=\alpha_{1}$ could be 
used for excluding $H_{1}$ when $p_{1}$ is small (less than 10\% in the figure).  
If the observed value of the test statistic $T$ was such that $(p_{0}, p_{1})$ 
was in the large rectangle, the data would be consistent with both hypotheses, 
while for the small rectangle near the origin, it would be inconsistent with 
both.  An alternative procedure for excluding $H_{1}$ is based on 
$CL_{s} = p_{1}/(1 - p_{0})$, and requires $(p_{0}, p_{1})$ to be below the 
sloping line.
\label{fig:lines}}
\end{center}
\end{figure}

%%%%%%%%%%%%%%%%%%%%%%%%%%%%%%%%%%%%%%%%%%%%%%%%%%%%%%%%%%%%%%%%%%%%%%%%%%%%%%%%
Figure~\ref{fig:lines} contains a plot of $p_{0}$ versus $p_{1}$, with the 
regions for which the double test either rejects $H_{0}$ or 
fails to reject it; these depend solely on $p_{0}$.  In the diagram, the 
critical value $\alpha_{0}$ for $p_{0}$ is shown at 0.05; this value is chosen 
here for clear visibility on the plot, rather than as a realistic choice.

In particle physics, when we fail to reject $H_{0}$, we want to see further 
whether we can exclude $H_{1}$.  Although
not as exciting as discovery, exclusion can be useful from a theoretical point
of view and also for the purpose of planning the next measurement.
The most famous example is the Michelson-Morley experiment, which excluded any 
significant velocity of the earth with respect to the aether and led to the
demise of the aether theory.  In figure~\ref{fig:lines}, the region $p_{1}\le\alpha_{1}$ 
is used for excluding $H_{1}$. The critical value $\alpha_{1}$ is usually 
chosen to be larger than the $p_{0}$ cut-off $\alpha_{0}$; 0.05 is a typical 
value. In the figure $\alpha_{1}$ is shown at 0.10. 

If $p_{0}$ and $p_{1}$ fall in the large rectangle at the top right of the plot
($p_{0}>\alpha_{0}$ and $p_{1}>\alpha_{1}$), we claim neither discovery of $H_{1}$ 
nor its exclusion: this is the no-decision region.  The small rectangle near 
the origin corresponds to both $p$-values being below their cut-offs, and the 
data are unlikely under either hypothesis.  It could correspond to the new 
physics occurring, but at a lower than expected rate.

%%%%%%%%%%%%%%%%%%%%%%%%%%%%%%%%%%%%%%%%%%%%%%%%%%%%%%%%%%%%%%%%%%%%%%%%%%%%%%%%
\subsection{The \texorpdfstring{$\mathbf{CL_{s}}$}{CLs} criterion}
%%%%%%%%%%%%%%%%%%%%%%%%%%%%%%%%%%%%%%%%%%%%%%%%%%%%%%%%%%%%%%%%%%%%%%%%%%%%%%%%
An alternative approach for exclusion of $H_{1}$ is the $CL_{s}$ criterion\cite{Read}. 
Because exclusion levels are chosen to have modest values (say 95\%), there is 
substantial probability (5\%) that $H_{1}$ will be excluded even when the 
experiment has little sensitivity for distinguishing $H_{1}$ from $H_{0}$
(the situation shown in figure~\ref{fig:pdfs}(a)).  Although professional 
statisticians are not worried about this, in particle physics it is regarded as 
unsatisfactory.  To protect against this, instead of rejecting $H_{1}$ on the
basis of $p_{1}$ being small, a cut is made on 
\begin{equation}
CL_{s}\;\equiv\; \frac{p_{1}}{1-p_{0}}, 
\end{equation}
i.e. on the ratio of the left-hand tails of the $H_{0}$ and $H_{1}$ pdf's.  Thus 
if the pdf's are almost indistinguishable, the ratio will be close to unity, and 
$H_{1}$ will not be excluded.  In figure~\ref{fig:lines}, the region below the 
dashed line referred to as `$CL_{s}$' shows where $H_{1}$ would be excluded.  
This is to be compared to the larger region below the horizontal line for the 
more conventional exclusion based on $p_{1}$ alone.  The $CL_{s}$ approach can 
thus be regarded as a conservative modification of the exact frequentist method; 
conservatism is the price to pay for the protection $CL_{s}$ provides against 
exclusion when there is little or no sensitivity to $H_{1}$.

%%%%%%%%%%%%%%%%%%%%%%%%%%%%%%%%%%%%%%%%%%%%%%%%%%%%%%%%%%%%%%%%%%%%%%%%%%%%%%%%
\subsection{Upper limits}
%%%%%%%%%%%%%%%%%%%%%%%%%%%%%%%%%%%%%%%%%%%%%%%%%%%%%%%%%%%%%%%%%%%%%%%%%%%%%%%%
As pointed out in the introduction, the pdf of $H_{1}$ often contains one or 
more parameters of interest whose values are not specified (e.g. the mass
of a new particle, the cross section of a new process, etc.).  It is then 
useful to determine the subset of $H_{1}$ parameter space where, with 
significance threshold $\alpha_{1}$, each parameter value is excluded by the 
observations.  In the frequentist paradigm, the complement of this subset is a 
$\textrm{CL}=1-\alpha_{1}$ confidence region.
  
For a simple and common example consider the case where the pdf of the data 
depends on the cross section $\mu$ of a new physics process: then $\mu>0$ if 
the process is present in the data ($H_{1}$ true), and $\mu=0$ otherwise 
($H_{0}$ true).  Suppose that the test statistic $T$ is stochastically 
increasing with $\mu$, meaning that for fixed $T$, increasing $\mu$ reduces the 
$p$-value $p_{1}$.  Then the set of $\mu$ values that {\em cannot} be excluded 
by the observations has an upper limit, and that upper limit has confidence 
level $1-\alpha_{1}$.

If instead of rejecting $H_{1}$ with the standard frequentist criterion 
$p_{1}\le \alpha_{1}$, we use the $CL_{s}$ criterion $CL_{s}\le\alpha_{1}$,
the above procedure yields a $CL_{s}$ upper limit for $\mu$, which is 
higher (i.e. weaker) than the standard frequentist upper limit.

In the previous example suppose that, instead of a cross section, $\mu$ is a 
location parameter for the test statistic $t$.  More precisely, suppose that 
the pdf of $t$ is of the form $f(t-\mu)$, with $f$ a continuous distribution.  
Then it can be shown that the 
%$\textrm{CL}=1-\alpha_{1}$ $CL_{s}$ upper limit 
upper  limit using $CL_s$ at the $1-\alpha_1$ level 
coincides exactly with the credibility $1-\alpha_{1}$ Bayesian upper limit 
obtained by assuming a uniform prior for $\mu$ under $H_{1}$ (i.e., a prior 
that is a non-zero constant for $\mu>0$, and zero elsewhere).  This result 
extends to the discrete case where $t$ is a Poisson-distributed event count 
with mean $\mu$ (see Appendix~\ref{AppA}).

%%%%%%%%%%%%%%%%%%%%%%%%%%%%%%%%%%%%%%%%%%%%%%%%%%%%%%%%%%%%%%%%%%%%%%%%%%%%%%%%
\subsection{Fixed-hypothesis contours in the 
\texorpdfstring{$\mathbf{(p_0,p_1)}$}{(p0,p1)} plane.}
\label{sec:FH_Contours}
\begin{figure}[p]
\begin{center}
\includegraphics[width=\textwidth]{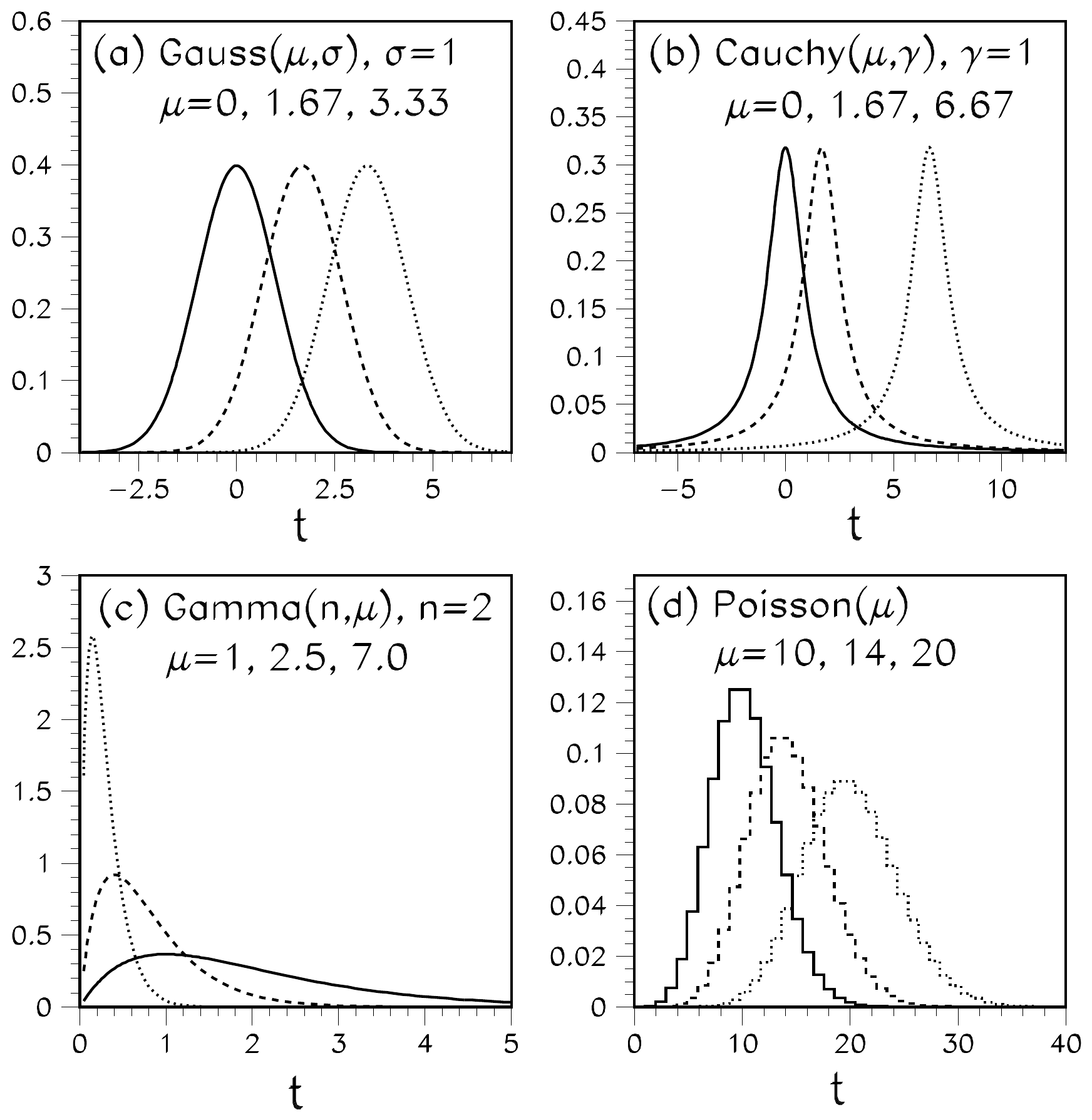}
\caption{Probability densities (plots a, b, and c) and probabilities (plot d)
 used to draw fixed-hypothesis contours on figure~\protect\ref{fig:fh_4examples}.
\label{fig:fourpdfs}}
\end{center}
\end{figure}

\begin{figure}[p]
\begin{center}
\includegraphics[width=\textwidth]{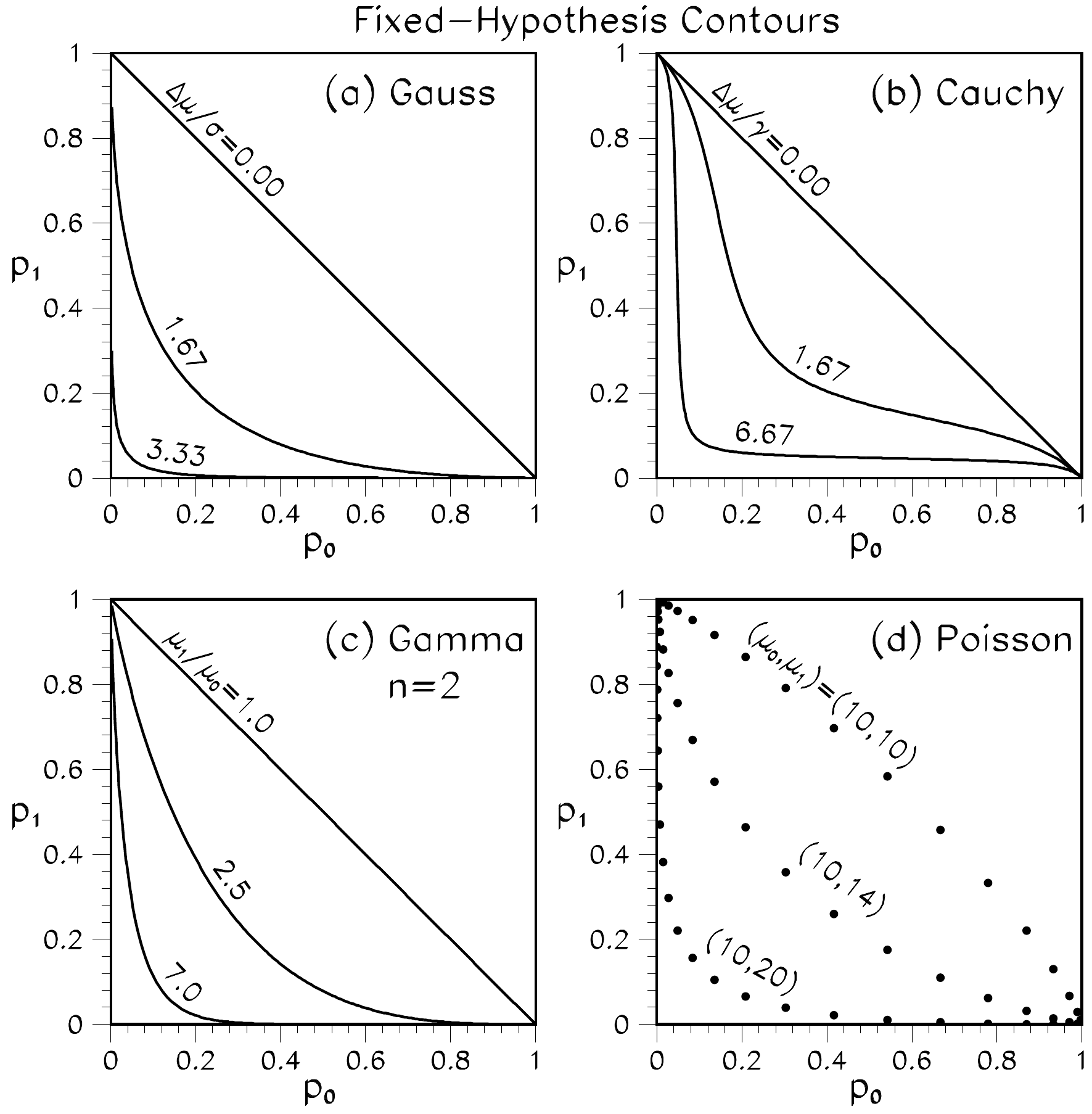}
\caption{Fixed-hypothesis contours in the $p_0$ versus $p_1$ plane for four
different choices of pdf: (a) Gauss, (b) Cauchy, (c) Gamma, and (d) Poisson.
In each case one is testing $H_{0}:\mu=\mu_{0}$ versus $H_{1}:\mu=\mu_{1}$
where $\mu$ is the mean of the distribution (or mode, in the case of Cauchy).
In the Gauss, Cauchy, and Gamma cases, the contours depend on a simple
combination of $\mu_{0}$ and $\mu_{1}$, but not on their individual values.
The line $p_{1}=1-p_{0}$ is an upper boundary for observable $p$-values in 
plots (a), (b), and (c), but not in the discrete Poisson case of plot (d).
\label{fig:fh_4examples}}
\end{center}
\end{figure}

%%%%%%%%%%%%%%%%%%%%%%%%%%%%%%%%%%%%%%%%%%%%%%%%%%%%%%%%%%%%%%%%%%%%%%%%%%%%%%%%
If we keep the hypotheses $H_{0}$ and $H_{1}$ fixed, but vary the observed data 
statistic $t$, the point $(p_{0,\textrm{obs}},p_{1,\textrm{obs}})$ will trace a 
contour in the $(p_{0},p_{1})$ plane.  In general this contour depends on the 
particular characteristics of each hypothesis, but useful simplifications may 
occur when the pdf of the test statistic is translation-invariant or enjoys 
other symmetries.  Here we give four examples based on the pdf's shown in
figure~\ref{fig:fourpdfs}.  The corresponding contours\footnote{Fixed-hypothesis 
contours on a $(p_0, p_1)$ plot are closely related to ROC (Receiver Operating 
Characteristic) curves, which have been used for many years in a variety of 
fields.} are drawn in figure~\ref{fig:fh_4examples} and assume that the test is 
of the basic form $H_{0}:\mu=\mu_{0}$ versus $H_{1}:\mu=\mu_{1}$, with 
$\mu_{1}>\mu_{0}$, and that the test statistic is $T$~\footnote{Following 
standard convention we write $T\sim f(t)$ to indicate that $f$ is the pdf of 
$T$ (use of the `$\sim$' symbol does not imply any kind of approximation).}.
The parameter $\mu$ could be related to the strength of a possible signal for a 
new particle with unknown mass. Increasing separation between $\mu_{0}$ and 
$\mu_{1}$ could then correspond to increasing amount of data; fixed amount of 
data and fixed particle mass, but increasing cross section; fixed amount of data 
and varying particle mass, with the cross section depending on the mass in a 
known way (i.e. raster scan).
\begin{description}
\item[Example 1:] $\mu$ is the mean of a Gaussian distribution of known width 
     $\sigma$:
     \begin{equation}
     T\sim \frac{e^{-\tfrac{1}{2}\bigl(\tfrac{t-\mu}{\sigma}\bigr)^{2}}}
                {\sqrt{2\pi}\sigma}.
     \end{equation}
     In this case the fixed-hypothesis contours only depend on $\Delta\mu/\sigma$,
     with $\Delta\mu\equiv\mu_{1}-\mu_{0}$, and have the form:
     \begin{equation}
     \textrm{erf}^{-1}(1-2\,p_{1}) + \textrm{erf}^{-1}(1-2\,p_{0}) = 
     \frac{\Delta\mu}{\sqrt{2}\,\sigma}.
     \end{equation}
     Figure~\ref{fig:fh_4examples}(a) shows three examples of this, with 
     $\Delta\mu/\sigma=0$ (when the locus is the diagonal line $p_0 + p_1 =1$), 
     $1.67$ and $3.33$.  As $\Delta\mu/\sigma$ increases, the curves pass closer 
     to the origin. 
%     $\blacksquare$
\item[Example 2:] $\mu$ is the mode of a Cauchy distribution with known half-width
     at half-height~$\gamma$:
     \begin{equation}
     T\sim \frac{\gamma}{\pi\,\left[\gamma^{2}+(t-\mu)^{2}\right]}.
     \end{equation}
     The contours have a simple expression that depends only on $\Delta\mu/\gamma$:
     \begin{equation}
     \tan\Bigl[\Bigl(1-2\,p_{1}\Bigr)\,\frac{\pi}{2}\Bigr] + 
     \tan\Bigl[\Bigl(1-2\,p_{0}\Bigr)\,\frac{\pi}{2}\Bigr] = \frac{\Delta\mu}{\gamma}.
     \end{equation}
     Example contours are shown in figure~\ref{fig:fh_4examples}(b).  
%     $\blacksquare$
\item[Example 3:] $\mu$ is an exponential decay rate:
     \begin{equation}
     T\sim \mu e^{-\mu t}
     \end{equation}
     Here the fixed-hypothesis contours depend only on the ratio of $\mu_{1}$ to
     $\mu_{0}$:
     \begin{equation}
     \ln(p_{1}) = \frac{\mu_{1}}{\mu_{0}} \ln(1-p_{0}).
     \end{equation}
     An interesting generalization is to perform the test on a combination of 
     $n$ independent decay time measurements $T_{i}$.  In this case the 
     likelihood ratio statistic is a one-to-one function of the sum of the 
     measurements, which we therefore take as our test statistic, 
     $T\equiv\sum_{i=1}^{n}T_{i}$.  The distribution of $T$ is Gamma$(n,\mu)$, 
     with $n$ the shape parameter and $\mu$ the rate parameter:
     \begin{equation}
     T\sim \frac{\mu^{n}\,t^{n-1}\,e^{-\mu t}}{\Gamma(n)};
     \end{equation}
     the fixed-hypothesis contours depend on $n$ and on the ratio
     $\mu_{1}/\mu_{0}$:
     \begin{equation}
     p_{1} = 1 - P\Bigl(n,\,\frac{1}{2}\frac{\mu_{1}}{\mu_{0}}\,\chi^{2}_{2n,p_{0}}\Bigr),
     \end{equation}
     where $P(n,z)$ is the regularized incomplete gamma function and 
     $\chi^{2}_{2n,p}$ is the $p$-quantile of a chisquared distribution 
     with $2n$ degrees of freedom.  Some example contours are shown in
     figure~\ref{fig:fh_4examples}(c).  Unlike Gaussian or Cauchy contours, 
     gamma contours are not symmetric around the main diagonal of the plot.
%     $\blacksquare$
\item[Example 4:] $\mu$ is a Poisson mean:
     \begin{equation}
     T \sim \frac{\mu^{t}}{t!}\,e^{-\mu}\quad \textrm{($t$ integer)}.
     \end{equation}
     In this case the contours are discrete and must be computed numerically.
     Their dependence on $\mu_{0}$ and $\mu_{1}$ does not simplify.  A few 
     examples are plotted in figure~\ref{fig:fh_4examples}(d).
%     $\blacksquare$
\end{description}

A common feature of examples 1-3 above is that the region of the plot above 
the diagonal $p_{0}+p_{1}=1$ is empty.  This is a general consequence of 
our definition of one-sided $p$-values, and of the fact, suggested by the
requirement $\mu_{1}>\mu_{0}$, that the bulk of the pdf $f_{1}(t)$ under 
$H_{1}$ lies to the right of the bulk of the pdf $f_{0}(t)$ under $H_{0}$.  
In other words, for any $t$, the area under $f_{0}$ and to the right of $t$ 
is smaller than the corresponding area under $f_{1}$:
\begin{equation}
\textrm{For all } t:\;\int_{t}^{\infty} f_{0}(u)\, du\;\le\;\int_{t}^{\infty} f_{1}(u)\, du.
\end{equation}
On the left-hand side one recognizes $p_{0}$ and on the right-hand side
$1-p_{1}$, whence the inequality $p_{0}+p_{1}\le 1$ follows.  However, this is 
only strictly true for continuous pdf's.  For discrete pdf's, $p_{0}$ and 
$p_{1}$ {\em both} include the finite probability of the observation,
so that it may happen that $p_{0}+p_{1} > 1$ when $f_{0}$ and $f_{1}$ are
very close to each other.  This is evident in figure~\ref{fig:fh_4examples}(d)
for the case $\mu_{0}=\mu_{1}=10$.

%%%%%%%%%%%%%%%%%%%%%%%%%%%%%%%%%%%%%%%%%%%%%%%%%%%%%%%%%%%%%%%%%%%%%%%%%%%%%%%%
\subsection{Asimov data sets}
%%%%%%%%%%%%%%%%%%%%%%%%%%%%%%%%%%%%%%%%%%%%%%%%%%%%%%%%%%%%%%%%%%%%%%%%%%%%%%%%
In a given data analysis problem, any data set (real or artificial) for which 
the parameter estimators yield the true values is called an Asimov data 
set~\cite{Cowan2011}.  By evaluating a test statistic on an Asimov data set one
usually obtains an approximation to the median of that test statistic, and the
corresponding $p$-value will be the median $p$-value under the assumed 
hypothesis.  Median $p$-values are used to characterize the sensitivity of an
experiment.

A simple example of the use of the fixed-hypothesis contours is that they map 
the abscissa $p_{0}=0.5$ onto the median value of $p_{1}$ under $H_{0}$, and
vice-versa, the value $p_{1}=0.5$ is mapped onto the median of $p_{0}$ under
$H_{1}$.  These medians can be directly read off the plot.  For the Gaussian 
case with $\Delta\mu/\sigma=0.0$, $1.67$, or $3.33$, the median $p_{1}$ under 
$H_{0}$ is $0.5$, $4.7\times 10^{-2}$, or $4.3\times 10^{-4}$, respectively.  
By symmetry of the Gaussian density, these values are also those of the median
$p_{0}$ under $H_{1}$.

By the invariance of probability statements under one-to-one transformations 
of random variables, the median $CL_{s}$ under $H_{0}$ can be obtained by 
plugging $p_{0}=1/2$ into the definition of $CL_{s}$.  This yields:
\begin{equation}
\textrm{Med}_{H_{0}}(CL_{s}) 
\;=\; \left.CL_{s}\right|_{p_{0}=1/2} 
\;=\; \left.\frac{p_{1}}{1-p_{0}}\right|_{p_{0}=1/2}
\;=\; 2\, \left.p_{1}\right|_{p_{0}=1/2}
\;=\; 2\, \textrm{Med}_{H_{0}}(p_{1}).
\end{equation} 
Assuming $H_{0}$ is true, the median $CL_{s}$ for testing $H_{1}$ equals twice 
the median $p_{1}$.

%%%%%%%%%%%%%%%%%%%%%%%%%%%%%%%%%%%%%%%%%%%%%%%%%%%%%%%%%%%%%%%%%%%%%%%%%%%%%%%%
\subsection{Punzi sensitivity}
%%%%%%%%%%%%%%%%%%%%%%%%%%%%%%%%%%%%%%%%%%%%%%%%%%%%%%%%%%%%%%%%%%%%%%%%%%%%%%%%
For large enough separation of the pdf's, the fixed-hypothesis contour will keep 
out of the no-decision region.  Punzi~\cite{Punzi} defines sensitivity as the 
expected signal strength required for there to be a probability of at least 
$1-\alpha_{1}$ for claiming a discovery with significance $\alpha_{0}$ (e.g. a 
probability of 95\% for discovery at the level of $2.87\times 10^{-7}$).  This 
has the advantage that above the sensitivity limit, the data are guaranteed to 
provide rejection of $H_{0}$ at the significance level $\alpha_{0}$, or 
exclusion of $H_{1}$ at the significance level $\alpha_{1}$, or both;
the data cannot fall in the no-decision region.  In figure~\ref{fig:contours}, 
the Punzi sensitivity corresponds to a pdf separation for which the 
$(p_{0},p_{1})$ contour (not drawn) passes through the intersection of the 
vertical dot-dashed and horizontal dashed lines.  In the following we refer to 
this intersection as the `Punzi point'.

%%%%%%%%%%%%%%%%%%%%%%%%%%%%%%%%%%%%%%%%%%%%%%%%%%%%%%%%%%%%%%%%%%%%%%%%%%%%%%%%
\subsection{Effect of one-to-one transformations of the test statistic}
\label{sec:1to1transformations}
\begin{figure}[p]
\begin{center}
\includegraphics[width=\textwidth]{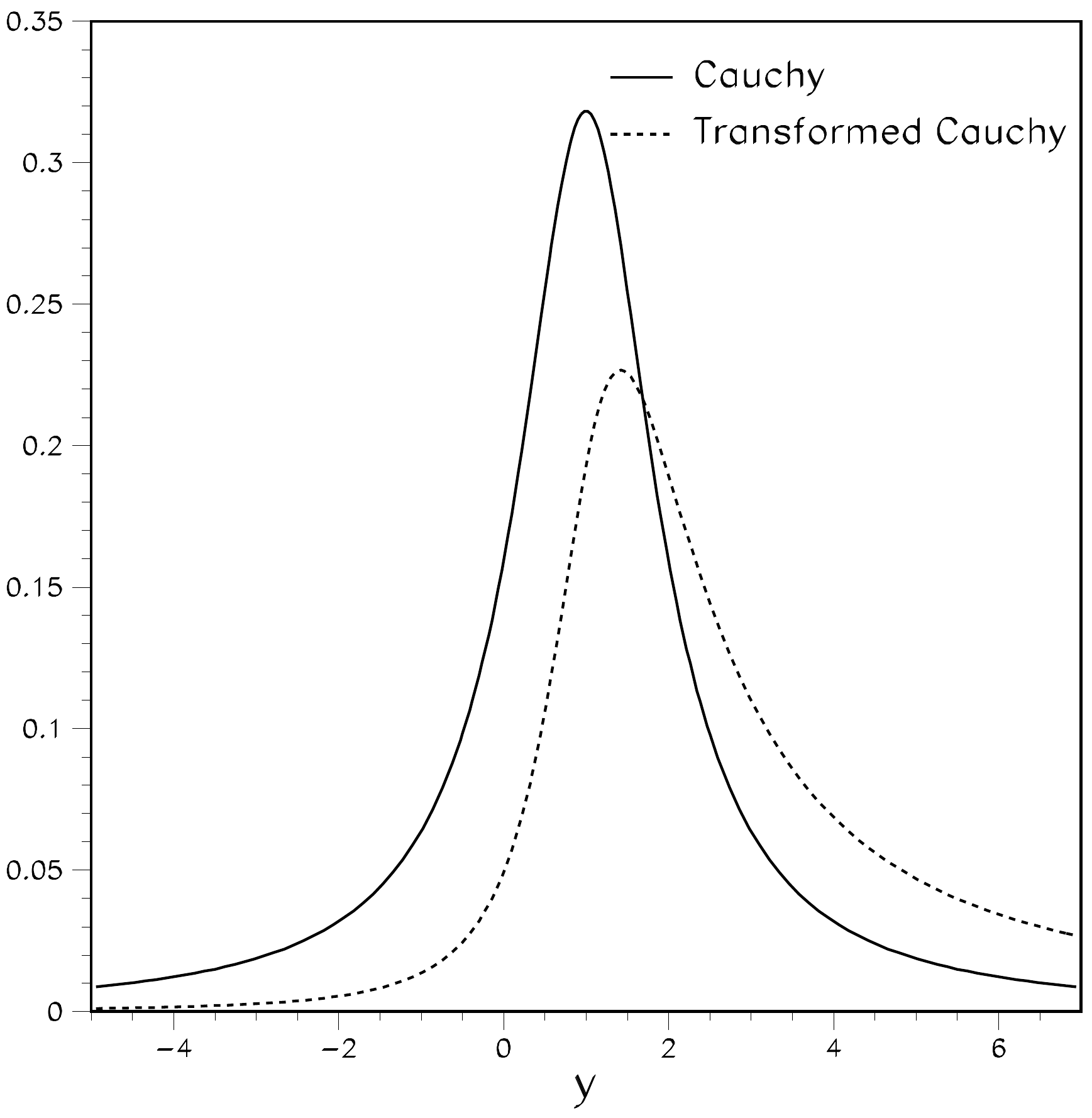}
\caption{Solid line: Cauchy probability density function with mode $\mu_{c}$ 
and scale parameter $\gamma$ both equal to 1.  Dashed line: asymmetric pdf 
obtained by a one-to-one transformation of the Cauchy density (see 
equation~\protect\eqref{eq:nonCauchy} in the text, with 
$\mu_{c}=\gamma=\Delta\mu/\sigma=1$).
\label{fig:cauchy}}
\end{center}
\end{figure}

%%%%%%%%%%%%%%%%%%%%%%%%%%%%%%%%%%%%%%%%%%%%%%%%%%%%%%%%%%%%%%%%%%%%%%%%%%%%%%%%
$P$-values are probabilities and therefore remain invariant under one-to-one
transformations of the test statistic on which they are based.  Plots of 
$p_{0}$ versus $p_{1}$ are similarly unaffected, but one must remember that 
these plots involve two hypotheses, and that effects of a transformation on
the pdf's of the test statistic under $H_{0}$ and $H_{1}$ are different.
This is the reason that, for example, the $p_{0}$ versus $p_{1}$ plot for 
testing the mode of a Gaussian pdf is not identical to the plot for testing 
the mode of a Cauchy pdf, even though Gaussian and Cauchy variates are related 
by one-to-one transformations (see examples~1 and~2 in 
section~\ref{sec:FH_Contours}).  We take a closer look at this particular case 
here.  Suppose that under $H_{0}$ ($H_{1}$) the test statistic $X$ is Gaussian 
with mean $\mu_{0}$ ($\mu_{1}$) and width $\sigma$.  Then, if $H_{0}$ is true, 
the transformation
\begin{equation}
X\longrightarrow Y\equiv \mu_{c} + \gamma\,\tan\left[
\frac{\pi}{2}\,\textrm{erf}\left(\frac{X-\mu_{0}}{\sqrt{2}\,\sigma}\right)\right]
\end{equation}  
maps $X$ into a Cauchy variate $Y$ with mode $\mu_{c}$ and scale parameter 
$\gamma$.  If on the other hand $H_{1}$ is true, $X$ will be mapped into a 
variate $Y$ with pdf
\begin{equation}
f(y) \;=\; \frac{\exp\left[-\frac{1}{2}\left(\frac{\Delta\mu}{\sigma}\right)^{2}\,+\,
\frac{\sqrt{2}\,\Delta\mu}{\sigma}\:\textrm{erf}^{-1}\left[\frac{2}{\pi}\,
\arctan\left(\frac{y-\mu_{c}}{\gamma}\right)\right]\right]}
{\pi\gamma\bigl[1 + \bigl(\frac{y-\mu_{c}}{\gamma}\bigr)^{2}\,\bigr]},
\label{eq:nonCauchy}
\end{equation}
which is an asymmetric density that depends on three parameters: $\mu_{c}$, 
$\gamma$, and $\Delta\mu/\sigma\equiv\lvert\mu_{1}-\mu_{0}\rvert/\sigma$; it 
reduces to a Cauchy density in the limit $\Delta\mu/\sigma\rightarrow 0$.
Figure~\ref{fig:cauchy} compares the two pdf's.  Thus, whereas in 
$X$ space we are testing a Gaussian hypothesis against a Gaussian hypothesis
with a different mean, in $Y$ space we are testing a Cauchy hypothesis 
against a hypothesis with a rather different, asymmetrical distribution.  However
the $p_{0}$ versus $p_{1}$ plot is the same in both spaces.

Another interesting property of one-to-one transformations of test statistics 
is that they preserve the likelihood ratio (since the Jacobian of the 
transformation cancels in the ratio).  Thus, if $f_{i}$ is the pdf of $X$ under
$H_{i}$, $i=0,1$, and we transform $X$ into $Y$ with pdf's $g_{i}$, we have:
\begin{equation}
\frac{g_{0}(y)}{g_{1}(y)} \;=\; \frac{f_{0}(x)}{f_{1}(x)}.
\end{equation}
Suppose now that the $X\longrightarrow Y$ transformation is the likelihood 
ratio transformation: $y \equiv f_{0}(x)/f_{1}(x)$.  Then it follows from the
above equation that
\begin{equation}
g_{0}(y) \;=\; y\, g_{1}(y),
\label{eq:LRtransform}
\end{equation}
a useful simplification.  If one prefers to work with the logarithm of the 
likelihood ratio, $q\equiv \ln y$, and $h_{i}(q)$ is the pdf of $q$ under 
$H_{i}$, then one finds:
\begin{equation}
h_{0}(q) \;=\; e^{q}\, h_{1}(q).
\label{eq:logLRtransform}
\end{equation}
Suppose for example that $f_{i}(x)$ is Gaussian with mean $\mu_{i}$ and width 
$\sigma$.  The pdf's of the log-likelihood ratio are then:
\begin{align}
h_{0}(q) & \;=\; \frac{1}{\sqrt{2\pi}\,\Delta\mu/\sigma}\;
             \exp\left[-\frac{1}{2}\left(\frac{q - \frac{1}{2}
             \left(\Delta\mu/\sigma\right)^{2}}{\Delta\mu/\sigma}\right)^{2}
             \:\right],\label{eq:logLRpdf0}\\[2mm]
h_{1}(q) & \;=\; \frac{1}{\sqrt{2\pi}\,\Delta\mu/\sigma}\;
             \exp\left[-\frac{1}{2}\left(\frac{q + \frac{1}{2}
             \left(\Delta\mu/\sigma\right)^{2}}{\Delta\mu/\sigma}\right)^{2}
             \:\right],\label{eq:logLRpdf1}
\end{align}
and it is straightforward to verify equation~\eqref{eq:logLRtransform}.  

The Gauss-versus-Gauss likelihood ratio $f_{0}(x)/f_{1}(x)$ in the above example 
is invariant under translations and rescalings of the original pdf's (i.e. under 
addition of a common constant to $\mu_{0}$, $\mu_{1}$, and $x$; and under 
multiplication of $\mu_{0}$, $\mu_{1}$, $\sigma$, and $x$ by a common factor).  
These two invariances reduce the three numbers ($\mu_{0}$, $\mu_{1}$, and 
$\sigma$) required to specify the $f_{i}(x)$ to a single one ($\Delta\mu/\sigma$) 
for the $h_{i}(q)$.  Note that $\Delta\mu/\sigma$ is the ratio of the difference 
in means to the standard deviation for the pair $(f_{0},f_{1})$ as well as 
$(h_{0},h_{1})$.

More generally, since the likelihood ratio transformation $x\rightarrow q$
is one-to-one, fixed-hypothesis contours obtained from the $h_{i}(q)$ are 
identical to those obtained from the $f_{i}(x)$.  

%%%%%%%%%%%%%%%%%%%%%%%%%%%%%%%%%%%%%%%%%%%%%%%%%%%%%%%%%%%%%%%%%%%%%%%%%%%%%%%%
\section{Outcome probabilities and error rates}
\label{sec:ErrorRates}
\begin{figure}[p]
\begin{center}
\includegraphics[width=\textwidth]{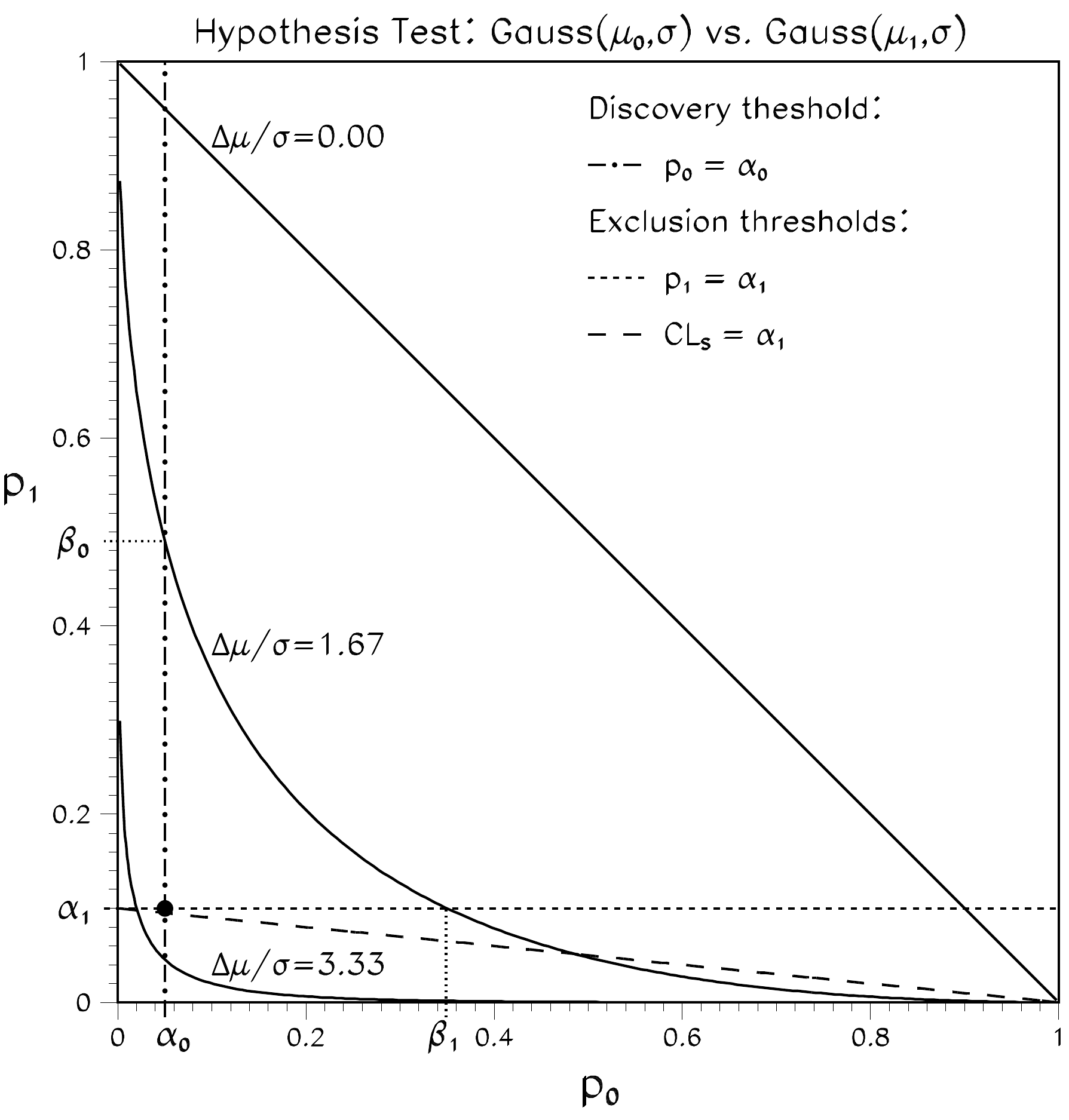}
\caption{Plot of $p_{0}$ versus $p_{1}$ with fixed-hypothesis contour lines, as 
in figure~\protect\ref{fig:fh_4examples}(a).  
For fixed pdf's under $H_{0}$ and $H_{1}$, the possible values of the test 
statistic $T$ correspond to a curve in the $(p_{0}, p_{1})$ plane.  The examples 
shown are for Gaussian pdf's where $\Delta\mu/\sigma$ is zero (i.e., identical 
pdf's under $H_{0}$ and $H_{1}$), 1.67 and 3.33.  When $\Delta\mu/\sigma = 3.33$, 
the separation of the pdf's is large enough that the data cannot fall in the 
large no-decision region, defined by $p_{0}>\alpha_{0}\;\&\;p_{1}>\alpha_{1}$. 
We refer to the black dot at the intersection of the horizontal dashed and 
vertical dot-dashed lines as the `Punzi point' (see text).  For a given choice
of $\alpha_{0}$, the intersection of the line $p_{0}=\alpha_{0}$ with the 
relevant contour has ordinate $\beta_{0}$, the probability of failing to reject
$H_{0}$ when $H_{1}$ is true (the plot shows this for the $\Delta\mu/\sigma = 1.67$ 
contour).  The relation between $\alpha_{1}$ and $\beta_{1}$ is similar.
\label{fig:contours}}
\end{center}
\end{figure}

%%%%%%%%%%%%%%%%%%%%%%%%%%%%%%%%%%%%%%%%%%%%%%%%%%%%%%%%%%%%%%%%%%%%%%%%%%%%%%%%
A useful feature of $(p_{0},p_{1})$ plots is that they help us map probabilities 
under $H_{0}$ to probabilities under $H_{1}$ and vice-versa, using a simple 
graphical method.  Suppose for instance that we are interested in the outcome 
$p_{0}\le 0.3$.  When $H_{0}$ is true this has probability 0.3, since $p_{0}$ 
is uniformly distributed under $H_{0}$.  To find the probability under $H_{1}$ 
we map the interval $0\le p_{0}\le 0.3$ onto the $p_{1}$ axis using the
appropriate contour on figure~\ref{fig:contours}, say the one with 
$\Delta\mu/\sigma=1.67$.  This yields the interval $0.13\le p_{1}\le 1$.  
Since $p_{1}$ is uniform under $H_{1}$, we can conclude that the outcome 
$p_{0}\le 0.3$ has probability $0.87$ under $H_{1}$.  In a similar way, it can 
be read from the figure that rejection of $H_{0}$ (the outcome 
$p_{0}\le\alpha_{0}$) has probability $1-\beta_{0}$ under $H_{1}$, where 
$\beta_{0}$ is the $p_{1}$ coordinate of the intersection of the line
$p_{0}=\alpha_{0}$ with the relevant contour.  As for rejection of $H_{1}$ 
(the outcome $p_{1}\le\alpha_{1}$), this has probability $1-\beta_{1}$ under 
$H_{0}$, where $\beta_{1}$ is the $p_{0}$ coordinate of the intersection of the
line $p_{1}=\alpha_{1}$ with the contour.

The graphical method allows one to derive the probabilities under $H_{0}$ and
$H_{1}$ of the four possible outcomes of the double test (see 
Table~\ref{tab:DoubleTestOutcomes}).  
\begin{table}[h]
\begin{center}
\begin{tabular}{|lclclcl|}
\hline
Double Test && \multirow{2}*{Decision} && Probability   && Probability     \\
Outcome     &&                         && Under $H_{0}$ && Under $H_{1}$   \\
\hline\hline
\multirow{2}*{$p_{0}\le\alpha_{0}\;\&\;p_{1}\le\alpha_{1}$} && Reject $H_{0}$ &&
\multirow{2}*{$\max(0,\alpha_{0}-\beta_{1})$} && 
\multirow{2}*{$\max(0,\alpha_{1}-\beta_{0})$} \\
 && Reject $H_{1}$ && && \\
\hline
\multirow{2}*{$p_{0}\le\alpha_{0}\;\&\;p_{1} > \alpha_{1}$} && Reject $H_{0}$ &&
\multirow{2}*{$\min(\alpha_{0},\beta_{1})$} && 
\multirow{2}*{$\min(1-\alpha_{1},1-\beta_{0})$} \\
 && Fail to reject $H_{1}$ && && \\
\hline
\multirow{2}*{$p_{0} > \alpha_{0}\;\&\;p_{1}\le\alpha_{1}$} && Fail to reject $H_{0}$ &&
\multirow{2}*{$\min(1-\alpha_{0},1-\beta_{1})$} && 
\multirow{2}*{$\min(\alpha_{1},\beta_{0})$} \\
 && Reject $H_{1}$ && && \\
\hline
\multirow{2}*{$p_{0} > \alpha_{0}\;\&\;p_{1} > \alpha_{1}$} && Fail to reject $H_{0}$ &&
\multirow{2}*{$\max(0,\beta_{1}-\alpha_{0})$} && 
\multirow{2}*{$\max(0,\beta_{0}-\alpha_{1})$} \\
 && Fail to reject $H_{1}$ && && \\
\hline
\end{tabular}
\caption{Possible outcomes of the double-test procedure, together with their
probabilities under $H_{0}$ and $H_{1}$.  As expected, the probabilities under
a given hypothesis all add up to one.
\label{tab:DoubleTestOutcomes}}
\end{center}
\end{table}
In computing these probabilities one needs
to handle separately the cases where the separation between the pdf's under 
$H_{0}$ and $H_{1}$ is smaller or larger than that corresponding to the Punzi 
sensitivity.  Consider for example the probability of outcome 
$p_{0} > \alpha_{0}\;\&\;p_{1} > \alpha_{1}$ under $H_{0}$.  Referring to 
Figure~\ref{fig:contours}, the contour with $\Delta\mu/\sigma=3.33$ passes below
the Punzi point, so that the desired probability is zero.  On the other hand, 
the contour with $\Delta\mu/\sigma=1.67$ passes above that point, and the segment 
of contour above both the $\alpha_{0}$ and $\alpha_{1}$ thresholds has 
probability $\beta_{1}-\alpha_{0}$ under $H_{0}$.  The probabilities for these 
two cases can be summarized as $\max(0,\beta_{1}-\alpha_{0})$, as shown in the
table.  An important caveat about the table is that the double test allows for
the possibility that an unspecified hypothesis other than $H_{0}$ and $H_{1}$ 
could be true, in which case a separate column of probabilities would be needed.
It is nevertheless reasonable to use this table for performance optimization 
purposes, since $H_{0}$ and $H_{1}$ are the two main hypotheses of interest.

Using Table~\ref{tab:DoubleTestOutcomes} one can compute various error rates as 
well as the power of the double test.  In analogy with the nomenclature of 
Neyman-Pearson tests, we can say that there are two Type-I errors, wrong 
decisions that are made when $H_{0}$ is true:
\begin{Displaylist}{XXXXXXXXXX}
\item[{\bf Type-Ia error:}]  Rejecting $H_{0}$ when $H_{0}$ is true.  The probability
     of this error is
     \begin{equation}
     \mathbb{P}(p_{0}\le\alpha_{0}\mid H_{0})\;=\; \max(0,\alpha_{0}-\beta_{1})
     + \min(\alpha_{0},\beta_{1})\;=\;\alpha_{0}.
     \end{equation}
     This is the Type-I error rate in a standard Neyman-Pearson test of $H_{0}$ 
     against $H_{1}$.
\item[{\bf Type-Ib error:}] Failing to reject $H_{1}$ when $H_{0}$ is true.  This has
     probability
     \begin{equation}
     \mathbb{P}(p_{1} > \alpha_{1}\mid H_{0})\;=\; \min(\alpha_{0},\beta_{1}) +
     \max(0,\beta_{1}-\alpha_{0})\;=\;\beta_{1}.
     \end{equation}
\end{Displaylist}
It is of course possible to commit both a Type-Ia and a Type-Ib error on the
same testing problem.  The rate of such double errors is not the product of
the individual rates $\alpha_{0}$ and $\beta_{1}$, but rather, as 
Table~\ref{tab:DoubleTestOutcomes} indicates, their minimum, 
$\min(\alpha_{0},\beta_{1})$.  Errors that are made when $H_{1}$ is true are 
called Type II:
\begin{Displaylist}{XXXXXXXXXX}
\item[{\bf Type-IIa error:}]  Rejecting $H_{1}$ when $H_{1}$ is true.  The 
     probability is
     \begin{equation}
     \mathbb{P}(p_{1}\le\alpha_{1}\mid H_{1})\;=\; \max(0,\alpha_{1}-\beta_{0})
     + \min(\alpha_{1},\beta_{0})\;=\;\alpha_{1}.
     \end{equation}
\item[{\bf Type-IIb error:}] Failing to reject $H_{0}$ when $H_{1}$ is true.  
     The rate of this error is
     \begin{equation}
     \mathbb{P}(p_{0} > \alpha_{0}\mid H_{1})\;=\; \min(\alpha_{1},\beta_{0}) +
     \max(0,\beta_{0}-\alpha_{1})\;=\;\beta_{0}.
     \end{equation}
     This is the Type-II error rate in a standard Neyman-Pearson test of $H_{0}$ 
     against $H_{1}$.
\end{Displaylist}
The rate for committing both Type-II errors simultaneously is 
$\min(\alpha_{1},\beta_{0})$.  Finally, there is a Type-III error, which has no
equivalent in the Neyman-Pearson setup:
\begin{Displaylist}{XXXXXXXXXX}
\item[{\bf Type-III error:}] Failing to reject $H_{0}$ and $H_{1}$ when a third, 
     unspecified hypothesis is true.  Without additional information about
     this third hypothesis it is not possible to calculate the Type-III error
     rate.
\end{Displaylist}

Since there is more than one Type-II error, there is some arbitrariness in the
definition of the power of the double test.  One possibility is to define it as
the probability of committing neither of the two Type-II errors, that is, as 
the probability of rejecting $H_{0}$ {\em and} failing to reject $H_{1}$, when 
$H_{1}$ is true:
\begin{equation}
\mathbb{P}(p_{0}\le\alpha_{0}\;\&\;p_{1}>\alpha_{1}\mid H_{1})\;=\;
1\,-\,\min(\alpha_{1},\beta_{0}) \;=\;
\min(1-\alpha_{1},1-\beta_{0}).
\label{eq:DoubleTestPower}
\end{equation}
This is different from the power of the Neyman-Pearson test, which is
$1-\beta_{0}$.  Equation~\eqref{eq:DoubleTestPower} has a simple 
interpretation if we look at it in terms of the separation between the $H_{0}$ 
and $H_{1}$ pdf's (see figure~\ref{fig:pdfs}).  At low separation, $\beta_{0}$ 
is large, and the power is dominated by our ability to reject $H_{0}$.  At high 
separation (figure~\ref{fig:pdfs}c), $\beta_{0}$ is low, and the power is 
limited by our willingness to accept $H_{1}$ (as opposed to a third, unspecified 
hypothesis).

Instead of using $p$-values to decide between hypotheses, one can use likelihood
ratios to evaluate the evidence against them.  In this case error rates are 
replaced by probabilities of misleading evidence.  The corresponding discussion 
can be found in Section~\ref{sec:misevidprob}.

%%%%%%%%%%%%%%%%%%%%%%%%%%%%%%%%%%%%%%%%%%%%%%%%%%%%%%%%%%%%%%%%%%%%%%%%%%%%%%%%
\section{Likelihood ratios}
\label{sec:LikelihoodRatios}
%%%%%%%%%%%%%%%%%%%%%%%%%%%%%%%%%%%%%%%%%%%%%%%%%%%%%%%%%%%%%%%%%%%%%%%%%%%%%%%%
Rather than using $p$-values for discriminating between hypotheses, it is 
possible to make use of a likelihood ratio\footnote{Note that a likelihood ratio 
can be used as a test statistic $T$ within a $p$-value method, or directly, 
without the calibration provided by $p$-values.  It is the latter case that we 
are considering in this section.};  this would also be the starting point for 
various Bayesian methods.  

%%%%%%%%%%%%%%%%%%%%%%%%%%%%%%%%%%%%%%%%%%%%%%%%%%%%%%%%%%%%%%%%%%%%%%%%%%%%%%%%
\subsection{Likelihood-ratio contours}
%%%%%%%%%%%%%%%%%%%%%%%%%%%%%%%%%%%%%%%%%%%%%%%%%%%%%%%%%%%%%%%%%%%%%%%%%%%%%%%%
It is instructive to plot contours of constant likelihood ratio 
$\lambda_{01}\equiv L_{0}/L_{1}$ on the $p_{0}$ versus $p_{1}$ plot.  This needs 
some thought however, since a likelihood ratio calculation requires three input 
numbers (the values $\mu_{0}$ and $\mu_{1}$ of the parameter $\mu$ under $H_{0}$ 
and $H_{1}$, and the observed value $t$ of the test statistic), whereas a point 
in the $(p_{0},p_{1})$ plane only yields two numbers.  Our approach here is the 
following: for a set of contours with given $\lambda_{01}$, we fix the null 
hypothesis $\mu_{0}$ in order to map $p_{0}$ to $t$, then solve the 
likelihood-ratio constraint $\lambda_{01}=L_{0}(t,\mu_{0})/L_{1}(t,\mu_{1})$ for 
$\mu_{1}$, and finally use $t$ and $\mu_{1}$ to obtain $p_{1}$.  In this way, 
both the likelihood ratio and the value of $\mu$ under $H_{0}$ are constant 
along our likelihood-ratio contours, but in general the value of $\mu$ under 
$H_{1}$ varies point by point.  

If the test statistic $t$ itself is the likelihood ratio, the above procedure
needs to be adjusted, since now the pdf's of $t$ under $H_{0}$ and $H_{1}$ depend
on both $\mu_{0}$ and $\mu_{1}$ (see for example equations~\eqref{eq:logLRpdf0}
and~\eqref{eq:logLRpdf1} in section~\ref{sec:1to1transformations}).  There is
no longer a likelihood-ratio constraint to solve.  Instead, for pre-specified
values of $\mu_{0}$ and $t\equiv\lambda_{01}$, one maps $p_{0}$ into $\mu_{1}$,
and substitutes $t$, $\mu_{0}$ and $\mu_{1}$ into the expression for $p_{1}$.

\begin{figure}[p]
\begin{center}
\includegraphics[width=\textwidth]{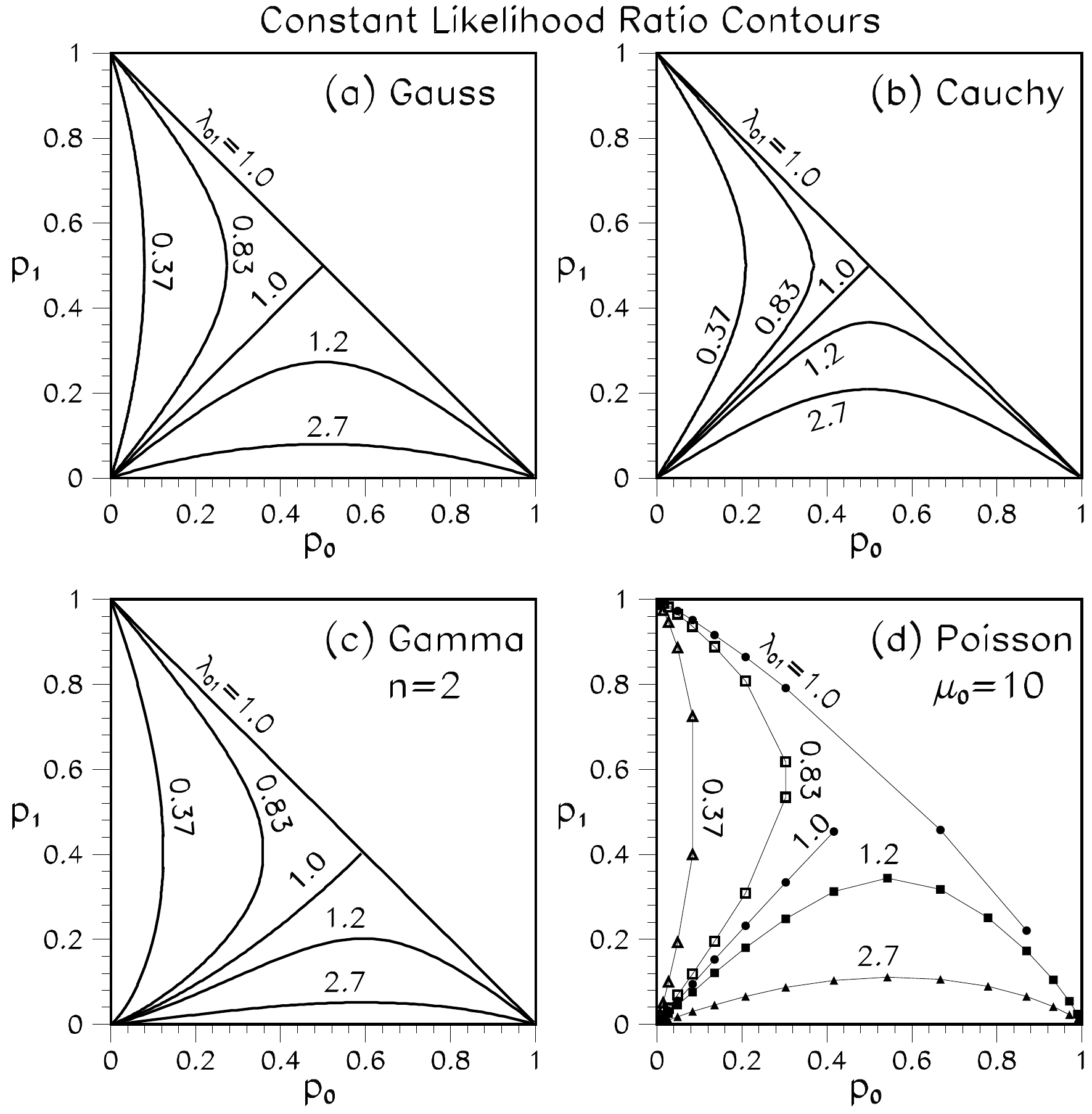}
\caption{Contours of constant likelihood ratio $\lambda_{01}\equiv L_{0}/L_{1}$ 
in the $p_{0}$ versus $p_{1}$ plane for four different choices of pdf: 
(a) Gauss, (b) Cauchy, (c) Gamma, and (d) Poisson, where the lines merely join 
up the discrete $(p_{0},p_{1})$ points as ``contours''.  In each case one is 
testing $H_{0}:\mu=\mu_{0}$ versus $H_{1}:\mu=\mu_{1}$ where $\mu$ is the pdf
mean for Gauss and Poisson, mode for Cauchy, and rate parameter for Gamma.
To facilitate comparison, the same contours are drawn in all four cases.  
In plot (d), points line up vertically across contours, since by construction 
$\mu_{0}$ is the same everywhere.
\label{fig:clr_4examples}}
\end{center}
\end{figure}

Remarkably, for some of the simple cases examined in section~\ref{sec:FH_Contours} 
it turns out that the likelihood-ratio contours are independent of $\mu_{0}$ and 
$\mu_{1}$.  The contours do depend on the family of pdf's to which the data are 
believed to belong, but not on the particular family members specified by the 
hypotheses.  For the examples of section~\ref{sec:FH_Contours}, the 
likelihood-ratio contours take the following forms:
\begin{description}
\item[Example 1:] $\mu$ is the mean of a Gaussian distribution of known width $\sigma$:
     \begin{equation}
     \left[\textrm{erf}^{-1}\left(1-2\,p_{1}\right)\right]^{2} - 
     \left[\textrm{erf}^{-1}\left(1-2\,p_{0}\right)\right]^{2} \;=\; \ln(\lambda_{01}).
     \end{equation}
\item[Example 2:] $\mu$ is the mode of a Cauchy distribution with known 
     half-width at half-height~$\gamma$:
     \begin{equation}
     \frac{1 + \left[\tan\bigl(\bigl(1-2\,p_{1}\bigr)\frac{\pi}{2}\bigr)\right]^{2}}
          {1 + \left[\tan\bigl(\bigl(1-2\,p_{0}\bigr)\frac{\pi}{2}\bigr)\right]^{2}}
     \;=\; \lambda_{01}.
     \end{equation}
\item[Example 3:] $\mu$ is an exponential decay rate:
     \begin{equation}
     \left[\frac{P^{-1}(n,p_{0})}{P^{-1}(n,1-p_{1})}\right]^{n}\;
     e^{P^{-1}(n,1-p_{1})-P^{-1}(n,p_{0})}\;=\;\lambda_{01},
     \end{equation}
     where $P^{-1}(n,x)$ is the inverse, with respect to the second argument, of 
     the regularized incomplete gamma function (i.e., $y=P^{-1}(n,x)$ is equivalent
     to $x=P(n,y)$).
\item[Example 4:] $\mu$ is a Poisson mean:\\
     There is no closed analytical expression, and the contours, which must be 
     computed numerically, depend on $\mu_{0}$ and $\mu_{1}$ (as opposed to just
     their difference or their ratio).
\end{description}
Figure~\ref{fig:clr_4examples} shows the $\lambda_{01}=0.37, 0.83, 1.0, 1.2$ and 
$2.7$ contours for these four cases.  Along the diagonal $p_{1}=1-p_{0}$ (or 
close to it in the Poisson case), the $H_{0}$ and $H_{1}$ pdf's are identical 
and $\lambda_{01}$ is unity.  For symmetric pdf's such as the Gaussian and 
Cauchy, the likelihood ratio is also unity along the other diagonal line, 
$p_{1}=p_{0}$.  This is because the observed value of the test statistic is then 
situated midway between the pdf peaks.  For asymmetric pdf's such as the gamma 
and Poisson the likelihood ratio is no longer unity when $p_{1}=p_{0}$, but 
there is still a $\lambda_{01}=1$ contour that starts at the origin of the plot 
and rises toward its middle.  Above and to the left of this curve, the 
likelihood ratio favors $H_{1}$; below it, $H_{0}$ is favored.

\begin{figure}[p]
\begin{center}
\includegraphics[width=\textwidth]{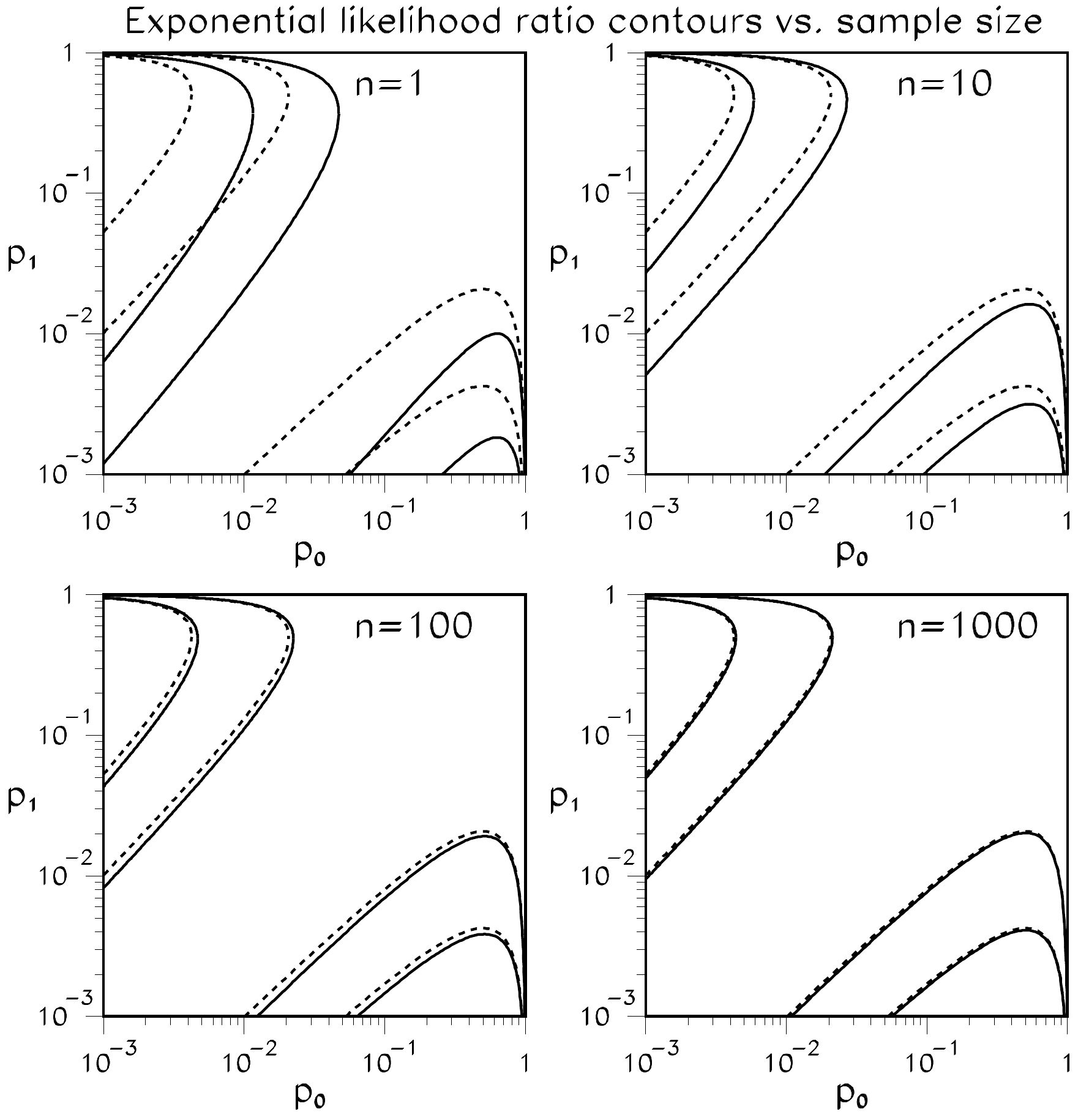}
\caption{Likelihood-ratio contours for testing the value of an exponential decay 
rate for four different values of the number $n$ of decay time measurements 
included in the test (solid lines), compared with the corresponding contours for 
a Gaussian test (dashed lines).  At large $n$, the exponential contours converge 
to the Gaussian ones.  From left to right in each plot, the contours correspond 
to $\lambda_{01}=1/32$, $1/8$, $8$, and $32$.
\label{fig:expvsgauss}}
\end{center}
\end{figure}

\begin{figure}[p]
\begin{center}
\includegraphics[width=\textwidth]{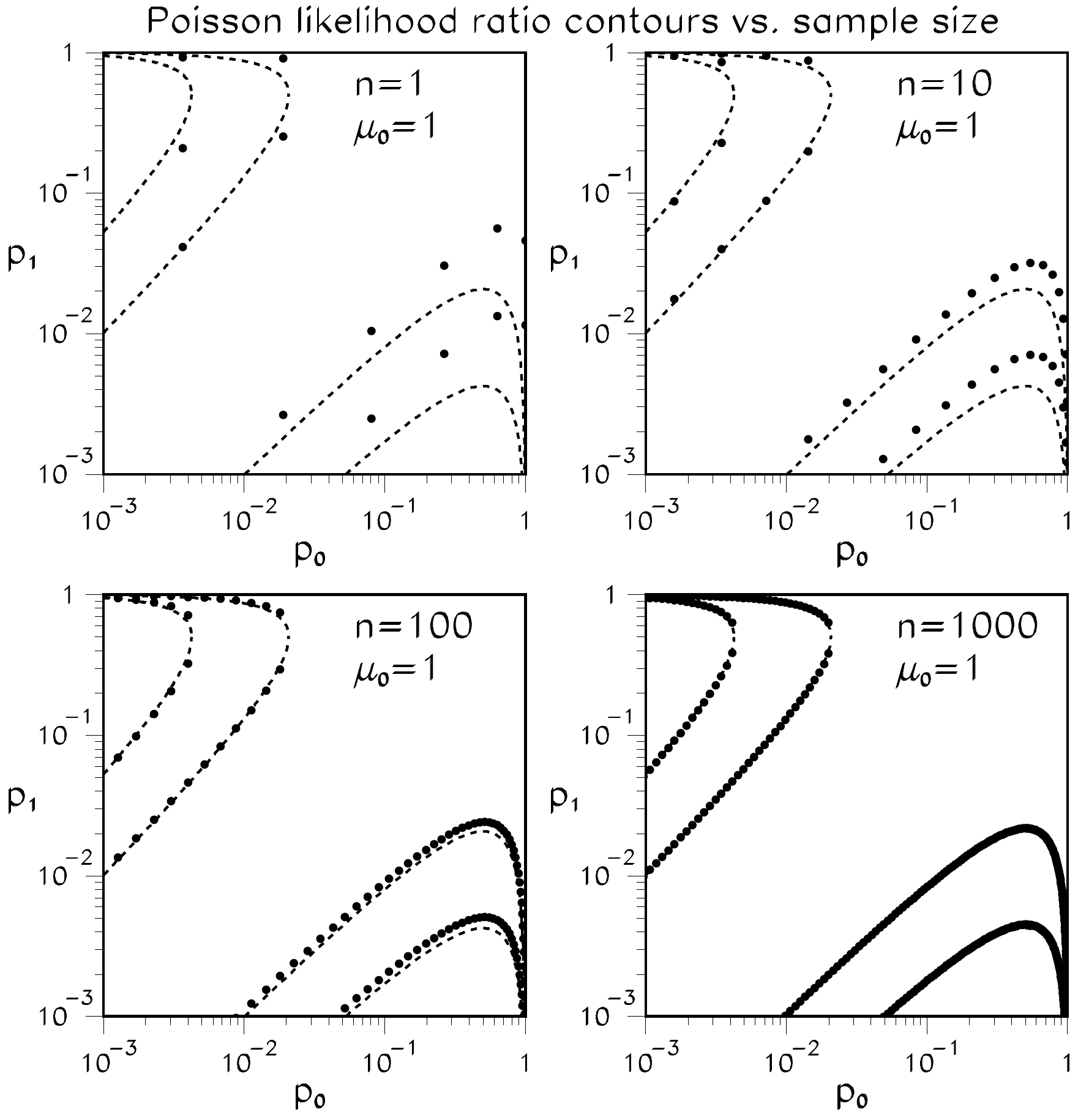}
\caption{Likelihood-ratio contours for testing the value of a Poisson mean for 
four different values of the number $n$ of measurements included in the test 
(dots), compared with the corresponding contours for a Gaussian test (dashed 
lines).  At large $n$, the Poisson contours converge to the Gaussian ones.  From 
left to right in each plot, the contours correspond to $\lambda_{01}=1/32$, 
$1/8$, $8$, and $32$.
\label{fig:poivsgauss}}
\end{center}
\end{figure}

Loosely stated, the central limit theorem asserts that the distribution of the
mean of $n$ measurements converges to a Gaussian as the sample size $n$ 
increases.  When the test statistic is defined as such a mean, likelihood ratio 
contours will converge to their shape for a Gauss versus Gauss test.  This is 
illustrated in figure~\ref{fig:expvsgauss} for the exponential/gamma case and in 
figure~\ref{fig:poivsgauss} for the Poisson case.

%%%%%%%%%%%%%%%%%%%%%%%%%%%%%%%%%%%%%%%%%%%%%%%%%%%%%%%%%%%%%%%%%%%%%%%%%%%%%%%%
\subsection{Comparison of \texorpdfstring{$\mathbf{p}$-values}{p-values} and 
likelihood ratios}
\label{sec:p_lr_comparison}
%%%%%%%%%%%%%%%%%%%%%%%%%%%%%%%%%%%%%%%%%%%%%%%%%%%%%%%%%%%%%%%%%%%%%%%%%%%%%%%%
A criticism against $p$-values is that they overstate the evidence against the null
hypothesis~\cite{Sellke2001,Berger2008}.  One aspect of this is that $p$-values 
tend to be impressively smaller than likelihood ratios.  The fact that they are 
not identical is no surprise. Likelihoods are calculated as the height of the 
relevant pdf at the observed value of the statistic $T$, while $p$-values use 
the corresponding tail area. Furthermore a $p$-value uses the pdf of a single
hypothesis, while a likelihood ratio requires the pdf's of two hypotheses.  
As can be seen from figure~\ref{fig:likelihood_ratio}, at constant $p_{0}$ 
(even if it is very small) $L_{0}/L_{1}$ can have a range of values, 
sometimes favoring $H_{1}$, sometimes $H_{0}$.  This will depend 
on the separation of the pdf peaks.  Thus for Gaussian pdf's, a $p_{0}$ value 
of $3\times 10^{-7}$ will favor $H_{1}$ provided $0<\Delta\mu/\sigma<10$, 
but for larger $\Delta\mu/\sigma$ the observed test statistic is closer to 
the $H_{0}$ peak than to $H_{1}$'s, and so even though the data are very 
inconsistent with $H_{0}$, the likelihood ratio still favors $H_{0}$ as 
compared with $H_{1}$.  

Another example is given in Table~\ref{tab:pvalLR}; this uses simple Poisson
hypotheses for both $H_{0}$ and $H_{1}$.  It involves a counting experiment 
where the null hypothesis $H_{0}$ predicts 1.0 event and the alternative $H_{1}$ 
predicts 10.0 events.  In a first run 10 events are observed; both $p_{0}$ and 
the likelihood ratio disfavor $H_{0}$.  Then the running time is increased by a 
factor of 10, so that the expected numbers according to $H_{0}$ and $H_{1}$ both 
increase by a factor of 10, to 10.0 and 100.0 respectively.  With 30 observed 
events, $p_{0}$ corresponds to about $5\sigma$ as in the first run, but despite 
this the likelihood ratio now strongly favors $H_{0}$.  This is simply because 
the $5\sigma$ $n_{obs}=10$ in the first run was exactly the expected value for 
$H_{1}$, but with much more data the  $5\sigma$ $n_{obs}=30$ is way below the 
$H_{1}$ expectation.  In fact, in the second run, the $p$-value approach rejects
both $H_{0}$ and $H_{1}$.

More data corresponds to increasing pdf separation.  Thus we are moving 
downwards on a line at constant $p_{0}$, resulting in a smaller $p_{1}$, and 
provided $p_{1}<1/2$, a larger $L_{0}/L_{1}$.  This is one motivation for 
hypothesis selection criteria that employ a decreasing value for the rejection 
threshold $\alpha_{0}$ as the amount of data increases.

It is interesting to contrast the exclusion regions for $H_{1}$ provided by 
cuts on $p_{1}$  and on the likelihood ratio $L_{0}/L_{1} = \lambda_{01}$ (see 
figures~\ref{fig:lines} and~\ref{fig:clr_4examples} respectively).  The main 
differences are at small and at large $p_{0}$, where the excluded region extends 
up to $p_{1}=\alpha_{1}$ for $p_{1}$ cuts, but to much smaller $p_{1}$ values 
for cuts on the likelihood ratio.  At large $p_{0}$, the likelihood cuts resemble 
more those provided by $CL_{s}$ (see figure~\ref{fig:lines}).  At small $p_{0}$, 
the likelihood cuts correspond to the exclusion $p_{1}$ cut-off $\alpha_{1}$ 
effectively decreasing as the $H_{0}$ and $H_{1}$ pdf's become more separated 
(e.g., as the amount of data collected increases).

\begin{table}
\begin{center} 
\begin{tabular}{|c|cc|}
\hline
\rule{0pt}{4.5mm}          &   First data set        &     Second data set       \\ 
\hline\hline
\rule{0pt}{4.5mm}$H_0$     &  Poisson, $\mu=1$       &     Poisson, $\mu=10$     \\
\hline
\rule{0pt}{4.5mm}$H_1$     &  Poisson, $\mu=10$      &     Poisson, $\mu=100$    \\
\hline
\rule{0pt}{4.5mm}$n_{obs}$ &      10                 &          30               \\
\hline\hline
\rule{0pt}{4.5mm}$p_0$     &  $1.1 \times 10^{-7}$   &    $2.5 \times 10^{-7}$   \\
                           &     $5.2\sigma$         &      $5.0\sigma$          \\
\hline
\rule{0pt}{4.5mm}$p_{1}$   &  $0.58$                 &    $2.2\times 10^{-16}$   \\
                           &  $-0.2\sigma$           &      $8.1\sigma$          \\
\hline
\rule{0pt}{4.5mm}$L_0/L_1$ &  $8 \times 10^{-7}$     &     $1.2  \times 10^{+9}$ \\
                           &  Strongly favors $H_1$ &   Strongly favors $H_0$  \\
\hline
\end{tabular}
\caption{Comparing $p$-values and likelihood ratios
\label{tab:pvalLR}}
\end{center}
\end{table}

%%%%%%%%%%%%%%%%%%%%%%%%%%%%%%%%%%%%%%%%%%%%%%%%%%%%%%%%%%%%%%%%%%%%%%%%%%%%%%%%
\subsection{Probability of misleading evidence in likelihood ratio tests}
\label{sec:misevidprob}
\begin{figure}[p]
\begin{center}
\includegraphics[width=\textwidth]{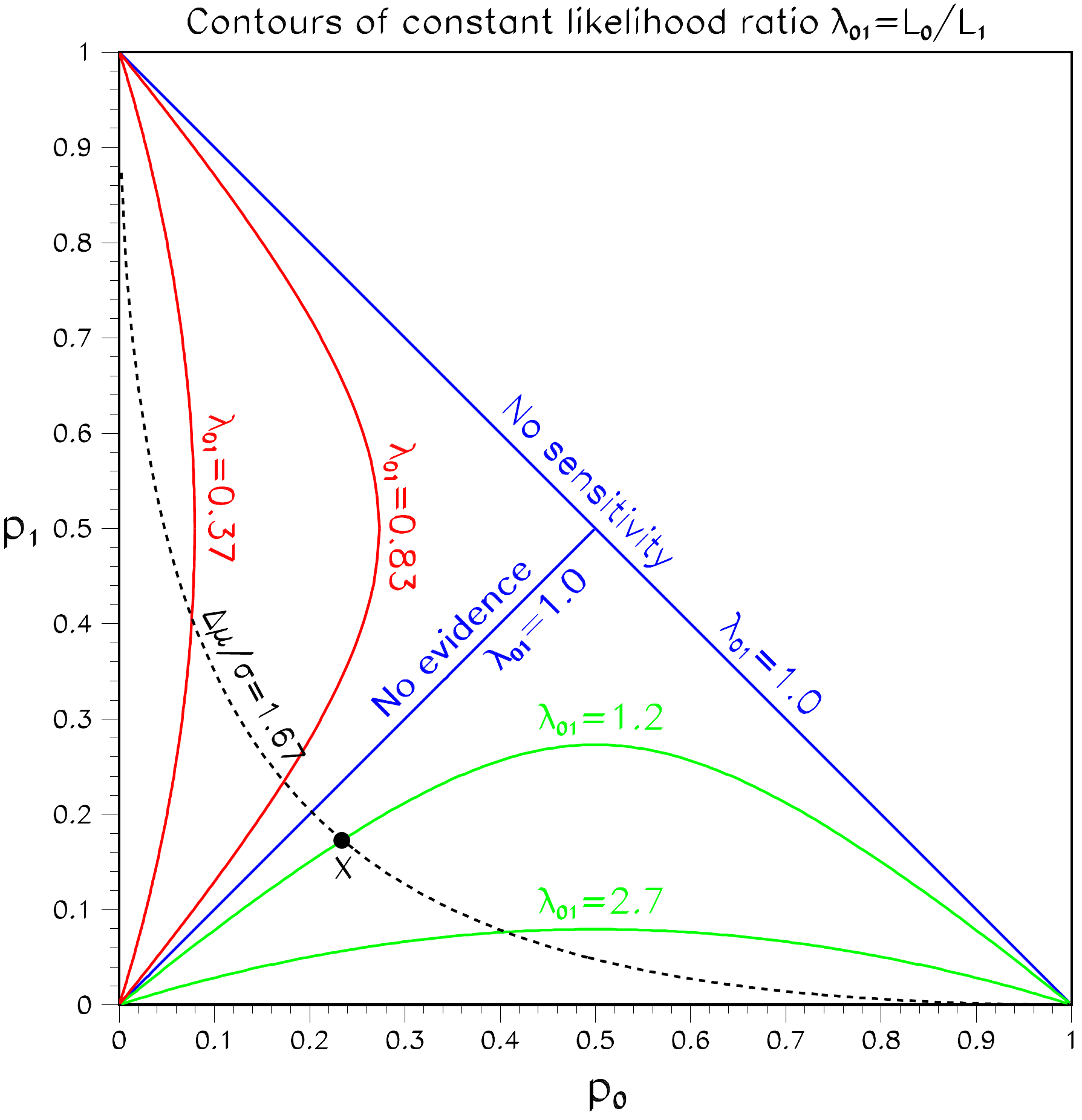}
\caption{Plot of $p_{0}$ versus $p_{1}$ with likelihood ratio contours (colored, 
solid lines), when the pdf's are Gaussians of equal width.  The likelihood ratio 
is unity along the $p_{1}=1-p_{0}$ diagonal, where $H_{1}$ is identical to 
$H_{0}$, and along the $p_{1}=p_{0}$ diagonal, where the observed value of the 
test statistic favors each hypothesis equally.  Going down a line of constant 
$p_{0}$ from $p_{1} = 0.5$ to $p_{1} = 0$ corresponds to increasing separation 
of the pdf's (e.g., more and more data), and also to increasing $L_{0}/L_{1}$.  
This gives rise to the possibility of $p_{0}$ being small while $L_{0}/L_{1}$ 
is large.  In fact one could exclude $H_{0}$ or $H_{1}$ based on the observed 
likelihood ratio instead of $p_{0}$ or $p_{1}$; the corresponding exclusion 
regions have a different shape from the $p$-value exclusion regions.
The dashed line is a contour of constant $\Delta\mu/\sigma$.  Its 
intersections with likelihood-ratio contours provides various probabilities of 
misleading evidence (see text).  
\label{fig:likelihood_ratio}}
\end{center}
\end{figure}

%%%%%%%%%%%%%%%%%%%%%%%%%%%%%%%%%%%%%%%%%%%%%%%%%%%%%%%%%%%%%%%%%%%%%%%%%%%%%%%%
When studying the evidence provided by the likelihood ratio $L_{0}/L_{1}$ in
favor of hypothesis $H_{0}$, an important quantity is the probability of 
misleading evidence.  This is defined by Royall~\cite{Royall2000} as the 
probability of observing $L_{0}/L_{1}>k$, for a given $k > 1$, when $H_{1}$ is 
true.  Figure~\ref{fig:likelihood_ratio} shows how this probability can be 
determined by drawing the appropriate fixed-hypothesis contour (dashed line, 
here corresponding to $\Delta\mu/\sigma=1.67$) on top of the likelihood-ratio 
contour of interest (here $L_{0}/L_{1}=1.2$).  Larger likelihood-ratio contours 
intersect the dashed line at lower values of $p_{1}$.  Therefore the probability 
of a larger likelihood ratio under $H_{1}$, i.e., the probability of misleading 
evidence, is given by the $p_{1}$-coordinate of the intersection point X. 

It is of course also possible to calculate the probability of misleading 
evidence that favors $H_{1}$ when $H_{0}$ is actually true.  For this
we look at the intersection of a fixed-hypothesis contour with a 
likelihood-ratio contour for which $L_{0}/L_{1} < 1$, and we are concerned 
about even smaller likelihood ratio values\footnote{Just as the cut-offs 
$\alpha_{0}$ and $\alpha_{1}$ for $p_{0}$ and $p_{1}$ are usually taken to be 
(very) different, similarly when using likelihood ratio cuts there is generally 
no necessity for one to be the reciprocal of the other.}.  The probability of 
misleading evidence is then given by the $p_{0}$-coordinate of that 
intersection.

Careful inspection of the shape of the likelihood-ratio contours in 
figure~\ref{fig:likelihood_ratio} reveals that the probabilities of misleading 
evidence are small at small values of $\Delta\mu/\sigma$ (where there is little 
chance of obtaining strong evidence in favor of either hypothesis), then 
increase to a maximum, and finally become small again at large 
$\Delta\mu/\sigma$.

\begin{figure}[H]
\begin{center}
\includegraphics[width=0.95\textwidth]{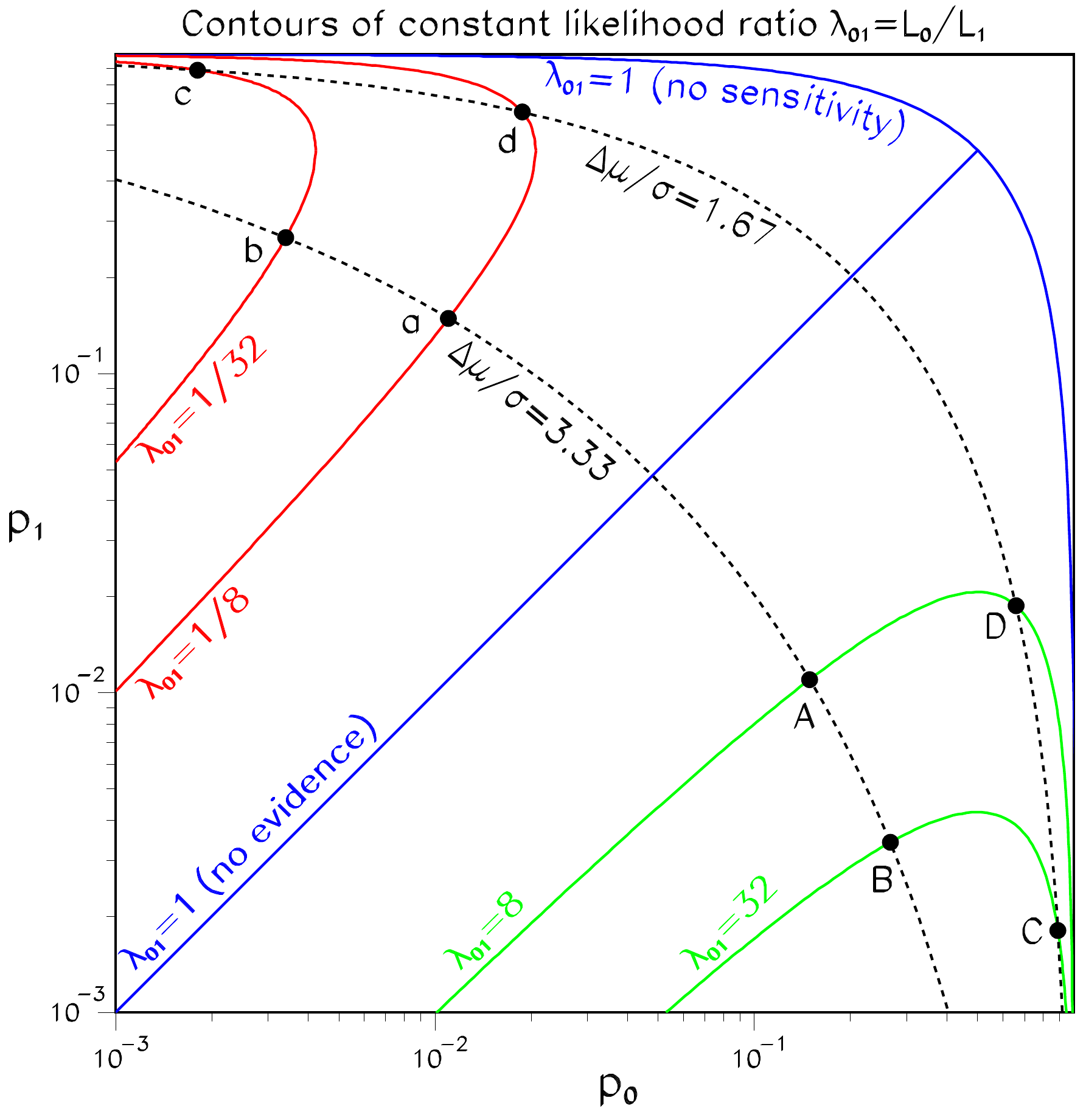}
\vspace*{-1mm}
\caption{
Log-log version of the plot of $p_{0}$ versus $p_{1}$, assuming that 
the pdf's of the test statistic under $H_{0}$ and $H_{1}$ are Gaussian with the
same width.  Solid lines show the likelihood-ratio contours for $L_{0}/L_{1}=1/32$, 
$1/8$, $1$, $8$ and $32$, and dashed lines show the fixed-hypothesis contours 
for $\Delta\mu/\sigma=1.67$ and $3.33$.  The $p_{0}$ coordinates of points $a$, 
$b$, $c$ and $d$, and the $p_{1}$ coordinates of points $A$, $B$, $C$ and $D$ 
yield the probabilities of misleading evidence listed in the table below.
\label{fig:likelihood_ratio_loglog}}
\end{center}
\end{figure}

The determination of probabilities of misleading evidence from $p_{0}$ and 
$p_{1}$ coordinates may give the impression that these probabilities could be
calculated from the observed likelihood ratio and reported `post-data'.  
According to the likelihoodist paradigm of statistics, this view is incorrect.  
As emphasized in ref.~\cite{Royall2000}, {\em all} the relevant evidence about 
the hypotheses is contained in the likelihood ratio.  The probabilities of 
misleading evidence should only be used for experiment-planning purposes, by
calculating them for standard likelihood ratio values. 
By convention, a value of $L_{0}/L_{1}=8$ is defined as `fairly strong' evidence 
in favor of $H_{0}$, whereas $L_{0}/L_{1}=32$ is said to be `strong' evidence.  
Likelihood-ratio contours for these values would not be visible on a linear plot 
such as figure~\ref{fig:likelihood_ratio}.  As shown in 
figure~\ref{fig:likelihood_ratio_loglog}, a log-log plot gives much better 
visualization.

%%%%%%%%%%%%%%%%%%%%%%%%%%%%%%%%%%%%%%%%%%%%%%%%%%%%%%%%%%%%%%%%%%%%%%%%%%%%%%%%
\section{Famous puzzles in statistics}
\label{sec:FamousPuzzles}
%%%%%%%%%%%%%%%%%%%%%%%%%%%%%%%%%%%%%%%%%%%%%%%%%%%%%%%%%%%%%%%%%%%%%%%%%%%%%%%%
The topic of $p$-values has generated many controversies in the statistics 
literature.  In this section we use $p_{0}$ versus $p_{1}$ plots to discuss a
couple of famous puzzles that initiated some of these controversies.

%%%%%%%%%%%%%%%%%%%%%%%%%%%%%%%%%%%%%%%%%%%%%%%%%%%%%%%%%%%%%%%%%%%%%%%%%%%%%%%%
\subsection{Sampling to a foregone conclusion}
\label{sec:foregone-conclusion}
\begin{figure}[p]
\begin{center}
\includegraphics[width=\textwidth]{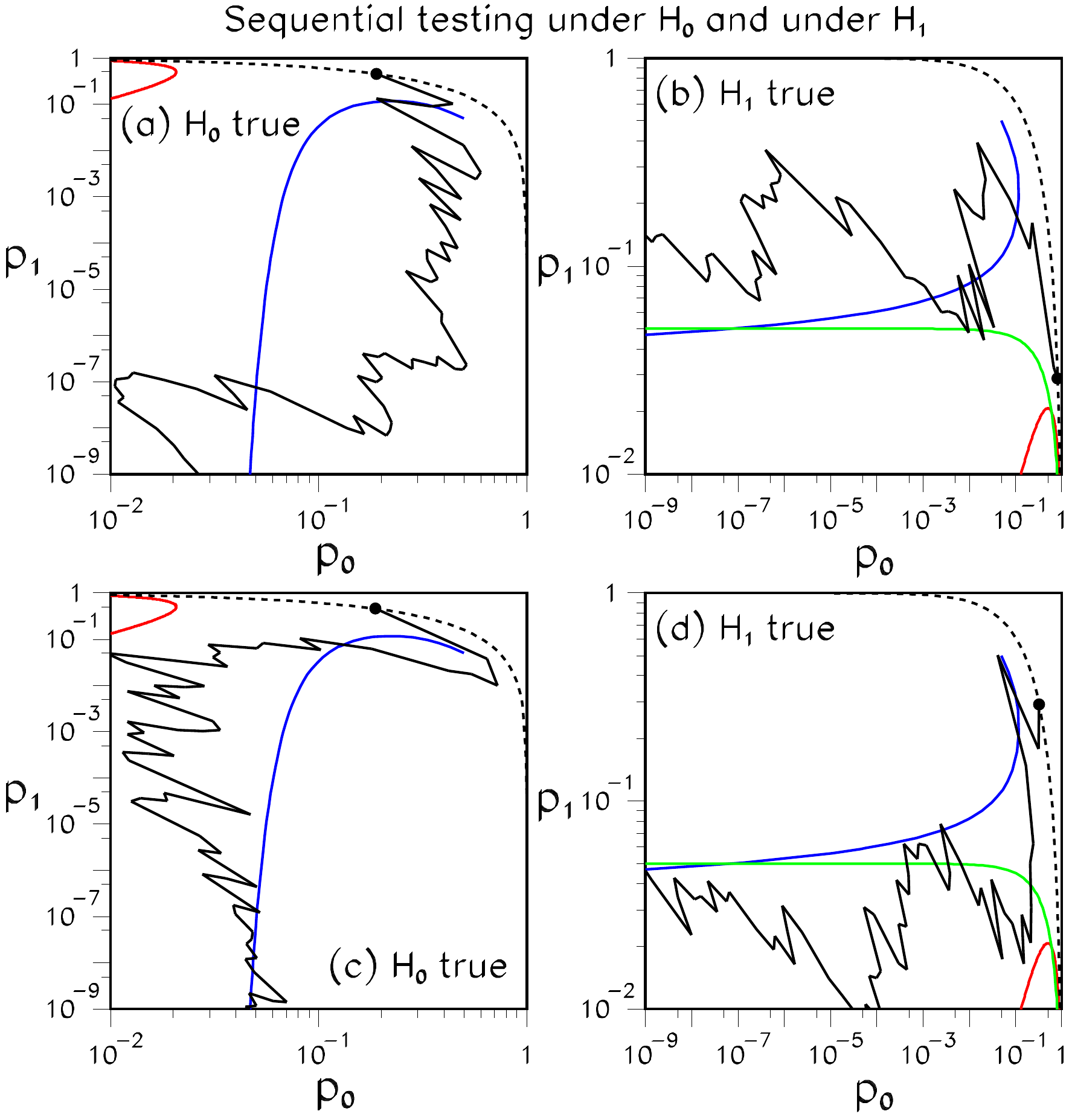}
\caption{Four examples of sequential testing on the mean of a Gaussian
distribution with unit width.  Plots (a) and (c) assume that $H_{0}$ is true,
whereas plots (b) and (d) assume the truth of $H_{1}$.  A sequential testing 
procedure (see text) describes a random walk in the $(p_{0},p_{1})$ plane 
(shown by the black broken lines).  The blue curves represent the boundary 
defined by the law of the iterated logarithm (LIL).  The red likelihood-ratio 
contours (for $\lambda_{01}=1/8$ in plots (a) and (c), and for $\lambda_{01}=8$ 
in plots (b) and (d)) are examples of decision boundaries that avoid the 
possibility of testing to a foregone conclusion implied by the LIL.  The green 
line in plots (b) and (d) represents the $CL_{s}=5\%$ decision boundary, which 
does not avoid this possibility.
\label{fig:seqtesting4x4}}
\end{center}
\end{figure}

%%%%%%%%%%%%%%%%%%%%%%%%%%%%%%%%%%%%%%%%%%%%%%%%%%%%%%%%%%%%%%%%%%%%%%%%%%%%%%%%
Suppose that in searching for a new physics phenomenon we adopt the following 
procedure:
\begin{enumerate}
\item Choose a discovery threshold $\alpha_{0}$, and let ${\cal E}$ be a set of
      candidate events, initially empty.
\item Add one candidate event to ${\cal E}$ and compute $p_{0}$, the $p$-value 
      to test the background-only hypothesis $H_{0}$ based on all events
      in ${\cal E}$.
\item If $p_{0}\le\alpha_{0}$, reject $H_{0}$, claim discovery, and stop;
      otherwise go back to step 2.
\end{enumerate}
If the new physics phenomenon can be modeled by a simple hypothesis $H_{1}$, we 
can also compute the $p$-value $p_{1}$ at step~2, and the whole procedure can be 
represented by a random walk in the $p_{0}$ versus $p_{1}$ plane.  At each step 
of the walk, the $p$-values are updated with the addition of a random new event.
Four examples of such random walks are shown in figure~\ref{fig:seqtesting4x4},
two assuming that $H_{0}$ is true, and two assuming that $H_{1}$ is true.

What is the chance of the above procedure stopping {\em when $H_{0}$ is true}?  
In other words, what is the probability of incorrectly claiming discovery with 
this procedure?  The answer, perhaps surprisingly, is 100\%, due to a result 
from probability theory known as the Law of the Iterated Logarithm (LIL).  The 
latter applies to any sequence of random variables $\{X_{1}, X_{2}, X_{3},\ldots\}$ 
that are independent and identically distributed with finite mean $\mu_{0}$ 
and variance $\sigma^{2}$.  Consider the $Z$-values constructed from partial 
sums of the $X_{i}$:
\begin{equation}
Z_{n}\;=\; \frac{\frac{1}{n}\sum_{i=1}^{n}X_{i}\,-\,\mu_{0}}{\sigma/\sqrt{n}},
\quad\textrm{for } n=1,2,3,\ldots
\end{equation}
The LIL states that with probability $100\%$ the inequality
\begin{equation}
\mid Z_{n}\mid \;\ge\; (1+\delta)\,\sqrt{2\ln\ln n}
\label{eqn:Zandn}
\end{equation}
holds for only finitely many values of $n$ when $\delta>0$ and for 
infinitely many values of $n$ when $\delta<0$.  At large $n$ the $Z_{n}$ will
be approximately standard normal and correspond to the $p$-values
\begin{equation}
p_{0}(n) \;=\; \int_{\mid Z_{n}\mid}^{\infty} \frac{e^{-t^{2}/2}}{\sqrt{2\pi}}\,dt
         \;=\; \frac{1}{2}\left[1 - 
               \textrm{erf}\left(\frac{\mid Z_{n}\mid}{\sqrt{2}}\right)\right],
\label{eqn:pandZ}
\end{equation}
so that the LIL of eqn.~\ref{eqn:Zandn} can be rephrased as stating that, 
as $n$ increases, the inequality
\begin{equation}
p_{0}(n) \;\le\; \frac{1}{2}\left[1-\textrm{erf}\left((1+\delta)\sqrt{\ln\ln n}\right)\right]
\label{eq:p0LIL}
\end{equation}
occurs infinitely many times if $\delta<0$.  In particular, regardless of how 
small $\alpha_{0}$ is, at large $n$ the right-hand side of~\eqref{eq:p0LIL} will 
become even smaller; therefore, if $\delta<0$ the LIL {\em guarantees} that 
$p_{0}(n)$ will cross the discovery threshold at some $n$, allowing the search 
procedure to stop with a discovery claim.  Crucial to this guarantee is the fact 
that inequality~\protect\eqref{eq:p0LIL} occurs {\em infinitely} many times for 
$\delta<0$; it will then {\em certainly} occur at $n$ large enough to force a
crossing of the discovery threshold.  In contrast, for $\delta>0$ there is a
value of $n$ beyond which there are no crossings (and there may indeed be none
at all for any $n$); rejection of $H_{0}$ is not guaranteed to occur.

\begin{figure}[p]
\begin{center}
\includegraphics[width=\textwidth]{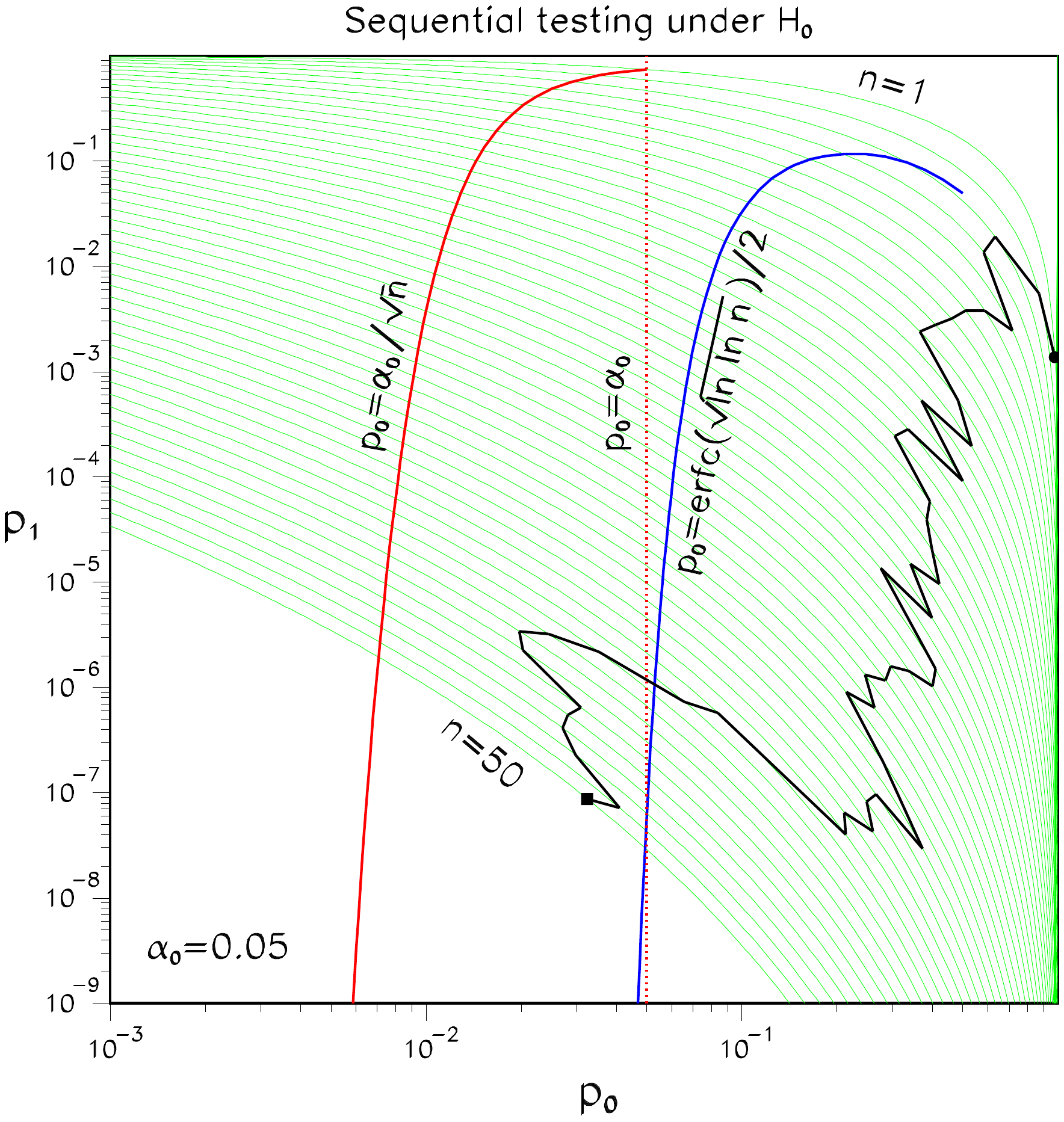}
\caption{Plot showing fifty fixed-hypothesis contours (green curves) crossed by 
a random walk (black broken line) associated with a sequential test procedure 
of the Gauss$(\mu_{0},\sigma)$ versus Gauss$(\mu_{1},\sigma)$ type.  At each 
step the sample size $n$ increases by one, and the walk moves to a contour with 
improved resolution.  Contours are labeled by the value of $n$.  The blue curve 
shows the relationship between $p_{0}$ and $n$ described by the LIL boundary.  
The red dotted line represents a fixed discovery threshold $\alpha_{0}$.  Since 
this line crosses to the large-$p_{0}$ side of the LIL boundary, it is guaranteed
to have the pathology of sampling to a foregone conclusion.  In contrast, with 
the $p_{0}$ cutoff set as $\alpha_{0}/\sqrt{n}$ (solid red line), repeated 
sampling does not necessarily lead to exclusion of a true $H_{0}$.
\label{fig:rootn_walk}}
\end{center}
\end{figure}

In terms of designing a coherent search procedure, one can view the LIL as 
defining an $n$-dependent boundary
\begin{equation}
\alpha_{\scriptscriptstyle\rm LIL}(n) \;=\; 
\frac{1}{2}\left[1-\textrm{erf}\left(\sqrt{\ln\ln n}\right)\right].
\end{equation}
Any discovery threshold with an $n$-dependence that causes it to exceed this
boundary at large $n$ is unsatisfactory since it is guaranteed to be crossed.  
It is instructive to draw the LIL boundary on a $p_{0}$ versus $p_{1}$ plot.  
To each value of $n$ there corresponds a fixed-hypothesis contour on the plot
(see figure~\ref{fig:rootn_walk}).  When testing $H_{0}$, one point on the LIL 
boundary is then given by the intersection of that contour with the line 
$p_{0}=\alpha_{\scriptscriptstyle\rm LIL}(n)$.  By connecting all such points 
across contours one obtains the blue lines drawn in figure~\ref{fig:rootn_walk}
and in figure~\ref{fig:seqtesting4x4}(a) and (c) (note that $n=2$ is the smallest 
integer for which $\alpha_{\scriptscriptstyle\rm LIL}(n)$ can be computed).  
When testing $H_{1}$, the LIL boundary is given by the intersections of the 
contours with the lines $p_{1}=\alpha_{\scriptscriptstyle\rm LIL}(n)$, as shown 
in figure~\ref{fig:seqtesting4x4}(b) and (d).

Focusing on plots (a) and (c) of figure~\ref{fig:seqtesting4x4}, we note that
when $H_{0}$ is true, the $p_{1}$ coordinate of random walks tends to decrease 
very rapidly as a function of $n$.  The $p_{0}$ coordinate is more stable, but
it does exhibit occasional excursions towards low $p_{0}$ values.  The LIL 
states that the number of such excursions to the left of the blue line is 
finite (not infinite) as $n$ goes to infinity.  However, any threshold curve 
to the right of the blue line will be crossed infinitely many times.  A constant 
threshold of the form $p_{0}=\alpha_{0}$ will be to the right of the blue line 
at large $n$ and is therefore unsatisfactory, in contrast with a threshold curve 
in the form of a likelihood ratio contour (see figure~\ref{fig:seqtesting4x4}) 
or with an $n$ dependence of the form $\alpha_{0}/\sqrt{n}$ (see 
figure~\ref{fig:rootn_walk}).

In particle physics we have constant thresholds of $3\sigma$ 
($\alpha_{0}=1.35\times 10^{-3}$) and $5\sigma$ 
($\alpha_{0}=2.87\times 10^{-7}$).  Due to the iteration of logarithms in the
LIL, it takes an enormously large value of $n$ for the blue line to cross these
thresholds, so that the problem is not practically relevant.  The statistician
I.~J.~Good once remarked that a statistician could ``cheat by claiming at a 
suitable point in a sequential experiment that he has a train to catch [\ldots]
But note that the iterated logarithm increases with fabulous slowness, so that 
this particular objection to the use of tail-area probabilities is theoretical 
rather than practical. To be reasonably sure of getting $3\sigma$ one would 
need to go sampling for billions of years, by which time there might not be any 
trains to catch.''~\cite{Good1965}

The LIL provides the weakest known constraint on the $n$-dependence of 
discovery thresholds.  It is a purely probabilistic characterization of tail
probabilities under a single hypothesis.  Much more stringent constraints can be
obtained by introducing an alternative hypothesis and using statistical 
arguments (see for example~\cite{BerryViele2008}).

%%%%%%%%%%%%%%%%%%%%%%%%%%%%%%%%%%%%%%%%%%%%%%%%%%%%%%%%%%%%%%%%%%%%%%%%%%%%%%%%
\subsection{The Jeffreys-Lindley paradox}
\label{sec:JL-paradox}
%%%%%%%%%%%%%%%%%%%%%%%%%%%%%%%%%%%%%%%%%%%%%%%%%%%%%%%%%%%%%%%%%%%%%%%%%%%%%%%%
The Jeffreys-Lindley paradox occurs in tests of a simple $H_{0}$ versus a
composite $H_{1}$, for example:
\begin{equation}
H_{0}:\; \mu=\mu_{0}\quad\textrm{ versus }\quad H_{1}:\; \mu>\mu_{0}.
\label{eq:JL-testing-problem-1}
\end{equation}
The paradox is that for some values of the observed test statistic $t$, the
value of $p_{0}$ can be small enough to cause rejection of $H_{0}$ while
the Bayes factor favors $H_{0}$.  Writing $L_{0}$, $L_{1}(\mu)$ for the 
likelihood under $H_{0}$, respectively $H_{1}$, the Bayes factor is defined by
\begin{equation}
B_{01}\;\equiv\;\frac{L_{0}}{\int\! L_{1}(\mu)\,\pi_{1}(\mu)\,d\mu},
\label{eq:JL-Bayes-factor}
\end{equation}
where $\pi_{1}(\mu)$ is a prior density for $\mu$ under $H_{1}$.  In order to 
understand the origin of the paradox, it helps to note that this Bayes factor 
can be rewritten as a weighted harmonic average of likelihood ratios:
\begin{equation}
B_{01}\;=\;\left[\int \frac{1}{\lambda_{01}(\mu)}\,\pi_{1}(\mu)\,d\mu \right]^{-1},
\label{eq:whmlr}
\end{equation}
with $\lambda_{01}(\mu)\equiv L_{0}/L_{1}(\mu)$.  As this formula suggests,
it will prove advantageous to look at the composite $H_{1}$ as a collection of 
simple hypotheses about the value of $\mu$, each with its own simple-to-simple 
likelihood ratio $\lambda_{01}(\mu)$ to the null hypothesis $H_{0}$.

In the following subsection we use this idea to develop basic insight into the 
origin of the Jeffreys-Lindley paradox.  Later subsections take a deeper
look at the conditions under which the paradox appears and at possible solutions.

%%%%%%%%%%%%%%%%%%%%%%%%%%%%%%%%%%%%%%%%%%%%%%%%%%%%%%%%%%%%%%%%%%%%%%%%%%%%%%%%
\subsubsection{Basic insight}
\label{sec:JL-Basic-Insight}
\begin{figure}[p]
\begin{center}
\includegraphics[width=\textwidth]{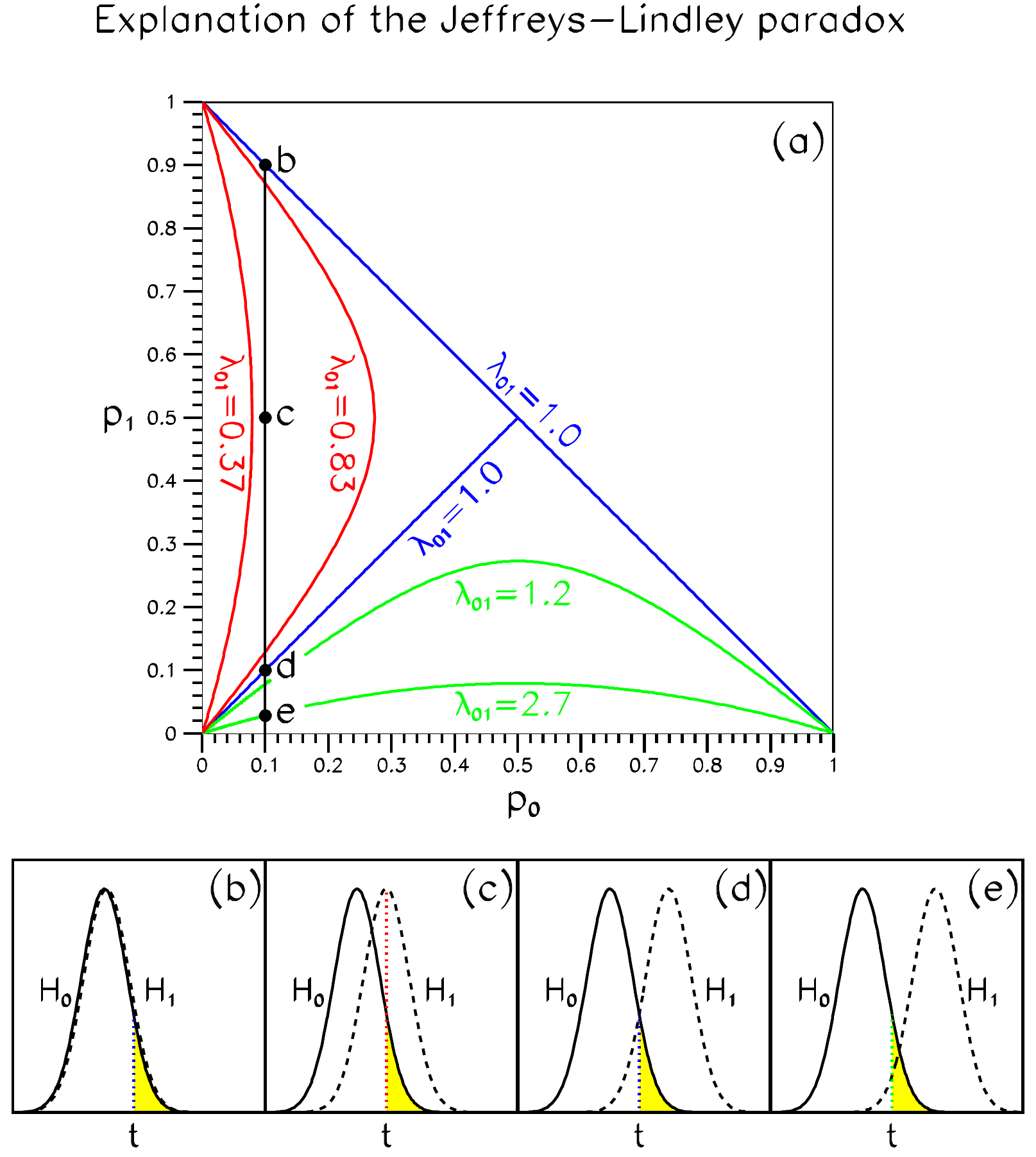}
\caption{Insight for the Jeffreys-Lindley paradox. The likelihod ratio contours 
in (a) are those of figure~\protect\ref{fig:clr_4examples}(a) for comparing 
hypotheses whose pdf's are equal-width Gaussians.  The line $bcde$ is at fixed 
$p_{0}$, with the points $b$ to $e$ corresponding to increasing separation of 
the $H_{0}$ and $H_{1}$ pdf's, as shown in diagrams (b) to (e) respectively.
\label{fig:jlillustration}}
\end{center}
\end{figure}

%%%%%%%%%%%%%%%%%%%%%%%%%%%%%%%%%%%%%%%%%%%%%%%%%%%%%%%%%%%%%%%%%%%%%%%%%%%%%%%%
Figure~\ref{fig:jlillustration} illustrates the paradox for the case where the
pdf of the test statistic $t$ is Gaussian with mean $\mu$ and standard deviation
$\sigma=1$.  As in Section~\ref{sec:p_lr_comparison}, consider a vertical 
line at the relevant $p_{0}$ in plot (a); this crosses a series of different 
$\lambda_{01}$ contours.  At point $b$, the $H_{0}$ and $H_{1}$ pdf's are 
identical (see plot (b)), and the likelihood ratio is unity.  Point $c$ is at 
$p_{1}=0.5$, with the $H_{1}$ pdf having its maximum exactly at the position of 
the data statistic $t$.  The likelihood ratio now favors $H_{1}$, and is in 
agreement with the small $p_{0}$ value in rejecting $H_{0}$.  Plots (d) and (e) 
show even larger separations between $H_{0}$ and $H_{1}$.  In plot (d), 
corresponding to point $d$ in plot (a), the position of the $H_{1}$ pdf is such 
that the data statistic $t$ is midway between the $H_{0}$ and $H_{1}$ peaks.  
Thus $p_{0}=p_{1}$ and point $d$ lies on the diagonal of plot (a), with the 
likelihood ratio again unity.  Finally, with the larger separation of plot (e), 
the likelihood ratio now favors $H_{0}$, even though $p_{0}$ is small; the 
likelihood ratio and $p_{0}$ lead to opposite conclusions.

To go from the series of simple $H_{1}$'s to the composite $H_{1}$ with 
unspecified $\mu$ in the Jeffreys-Lindley paradox, we take the weighted harmonic 
average of the likelihood ratios $\lambda_{01}$, with the weighting given by the 
prior $\pi_{1}(\mu)$ as in equation~\eqref{eq:whmlr}.  As we integrate along the 
vertical line in plot (a), the contributions between points $b$ and $d$ favor 
$H_{1}$.  Lower down, from $d$ to $e$ and beyond, $H_{0}$ is favored.  The 
Bayes factor will thus end up favoring $H_{0}$ if the integration range is wide 
enough\footnote{The value of $\mu$, which determines the separation between the 
corresponding simple $H_{1}$ and $H_{0}$, varies non-linearly with distance 
along the line $bcde$, such that there is generally a far wider range of $\mu$ 
values below the diagonal than above it.} and suitably weighted by the prior
$\pi_{1}(\mu)$.  This explains the mechanism by which the Jeffreys-Lindley 
paradox can occur.

%%%%%%%%%%%%%%%%%%%%%%%%%%%%%%%%%%%%%%%%%%%%%%%%%%%%%%%%%%%%%%%%%%%%%%%%%%%%%%%%
\subsubsection{Regions in the plane of \texorpdfstring{$\mathbf{p_{0}}$}{p0} 
               versus prior-predictive \texorpdfstring{$\mathbf{p_{1}}$}{p1}}
%%%%%%%%%%%%%%%%%%%%%%%%%%%%%%%%%%%%%%%%%%%%%%%%%%%%%%%%%%%%%%%%%%%%%%%%%%%%%%%%
To visualize the conditions under which the Jeffreys-Lindley paradox appears, we 
generalize the $p_{0}$ versus $p_{1}$ plot to the case of a composite $H_{1}$ by 
making use of the prior-predictive $p$-value~\cite{Box1980}; this is a 
prior-weighted average $p$-value over $H_{1}$:
\begin{equation}
p_{1pp}\;\equiv\; \int \! \pi_{1}(\mu) \int_{-\infty}^{t_{0}} f(t\mid\mu)\, dt\,d\mu,
\label{eq:prior-pred-p1}
\end{equation}
where $f(t\mid\mu)$ is the pdf of $T$ and $t_{0}$ its observed value.  To fix 
ideas, assume that $f$ is Gaussian with mean $\mu$ and standard deviation 
$\sigma$ (not necessarily equal to $1$), and that the prior $\pi_{1}(\mu)$ is the 
indicator function of the interval $[\mu_{0},\mu_{0}+\tau]$ for some positive 
$\tau$:
\begin{equation}
\pi_{1}(\mu) \;\equiv\; \pi_{1}(\mu\mid\tau)
             \;   =  \; \frac{1}{\tau}\; \mathds{1}_{[\mu_{0},\,\mu_{0}+\tau]}(\mu)
             \;   =  \; \begin{cases}
                         \dfrac{1}{\tau} & \textrm{ if}\quad \mu_{0} < \mu \le \mu_{0}+\tau,\\[3mm]
                         0               & \textrm{ otherwise.}
                        \end{cases}
\label{eq:H1-prior}
\end{equation}
In the absence of detailed prior information about $\mu$, one could think of 
this prior as modeling the range of $\mu$ values deemed to be theoretically 
and/or experimentally relevant for the search experiment of interest.  In any 
case the exact shape of $\pi_{1}$ is not material to the paradox, only the ratio 
of length scales $\tau/\sigma$ is.  Reference~\cite{Cousins2013} discusses the 
choice of $\tau$ in several particle physics experiments.

Use of $p_{1pp}$ calls for a couple of caveats.
First, a small value of $p_{1pp}$ does not imply that all values of $\mu$ under 
$H_{1}$ are disfavored.  In general it only provides evidence against the 
overall model (prior plus pdf) under $H_{1}$.  However with the particular 
choice of prior~\eqref{eq:H1-prior}, and assuming that $\tau/\sigma$ is 
sufficiently large, small $p_{1pp}$ implies that the vast majority of $\mu$ 
values under $H_{1}$ are unable to explain the data.  Second, the distribution 
of $p_{1pp}$ under a fixed value of $\mu$ in $H_{1}$ is not uniform.  Hence, in 
a linear plot of $p_{0}$ versus $p_{1pp}$, distances along the $p_{1pp}$ axis 
cannot be interpreted as probabilities under a fixed $\mu$ in $H_{1}$ (contrast 
Section~\ref{sec:ErrorRates}).  However, such distances can still be interpreted
as prior-predictive probabilities, with pdf given by the integral of $f(t\mid\mu)$
over $\pi_{1}(\mu)$.

\begin{figure}[p]
\begin{center}
\includegraphics[width=\textwidth]{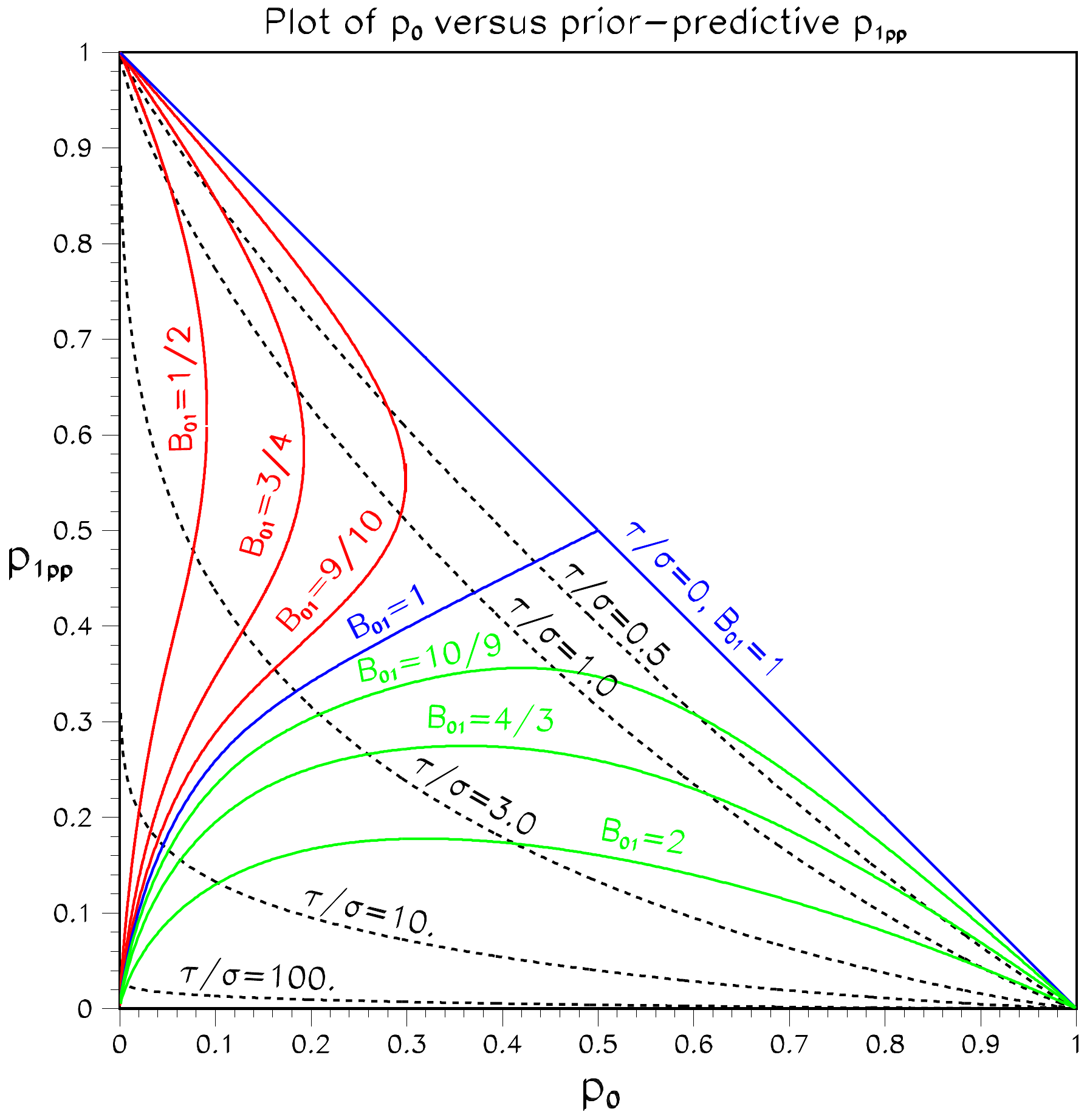}
\caption{
Plot of $p_{0}$ versus prior-predictive $p_{1}$ for testing $H_{0}: \mu=\mu_{0}$
versus $H_{1}: \mu>\mu_{0}$.  The test statistic has a Gaussian distribution 
with mean $\mu$ and standard deviation $\sigma$.  The prior for $\mu$ under 
$H_{1}$ equals $1/\tau$ for $\mu_{0} < \mu \le \mu_{0}+\tau$ and is zero
otherwise.  Fixed-hypothesis contours (dashed lines) are labeled by the value
of $\tau/\sigma$.  Constant Bayes factor contours (colored solid lines) are 
also shown.
\label{fig:ppp1_v_p0_lin}}
\end{center}
\end{figure}

Figure~\ref{fig:ppp1_v_p0_lin} shows fixed-hypothesis contours (fixed $\mu_{0}$,
$\sigma$, and $\tau$) and constant Bayes factor contours in the $p_{0}$ versus 
$p_{1pp}$ plane.  For the testing situation examined here, fixed-hypothesis 
contours only depend on the ratio $\tau/\sigma$ and are labeled accordingly.  
The constant Bayes factor contours are labeled by the value of $B_{01}$.  
For $\tau/\sigma=0$, $H_{1}$ coincides with $H_{0}$ and the resulting 
fixed-hypothesis contour is a subset of the $B_{01}=1$ contour.  
As $\tau/\sigma$ increases, the ability of the test to distinguish between 
$H_{0}$ and $H_{1}$ also increases.  Figure~\ref{fig:ppp1_v_p0_log} presents 
a log-log version of the same plot.  This allows the drawing of contours with a 
wider range of Bayes factor values, $B_{01} = 1$, $3$, $20$, and $150$.  
According to ref.~\cite{KassRaftery1995}, a Bayes factor between 1 and 3 
represents evidence ``not worth more than a bare mention;'' between 3 and 20, 
``positive;'' between 20 and 150, ``strong;'' and greater than 150, ``very 
strong.''  One can identify the following regions in the plot:
\begin{description}
\item[Upper Left:]  
      At small values of $p_{0}$ and small values of $\tau/\sigma$ (red contour
      region), the Bayes factors disfavor $H_{0}$.  There is agreement between
      Bayes factors and $p$-values.
\item[Lower Left:]
      At small values of $p_{0}$ and large values of $\tau/\sigma$ (green contour
      region), the Bayes factors favor $H_{0}$.  There is disagreement between
      Bayes factors and $p$-values.  This is where the Jeffreys-Lindley paradox
      shows up.  For a numerical example, consider a $p_{0}$ value of
      $2.87\times 10^{-7}$ ($5\sigma$); the corresponding Bayes factor in
      favor of $H_{0}$ will then be $1$, $3$, $20$, or $150$, if the ratio 
      $\tau/\sigma$ is approximately $6.7\times 10^{5}$, $2.0\times 10^{6}$,
      $1.3\times 10^{7}$, or $1.0\times 10^{8}$, respectively.  Note the 
      extremely large values of $\tau/\sigma$ required for producing the paradox.
      This is a consequence of the stringent $5\sigma$ convention applied to 
      discovery claims in particle physics.
\item[Upper Right:]
      This is a region with relatively large values of $p_{0}$ and $p_{1}$, and
      where the Bayes factor hovers around 1.  Regardless of how one looks at it,
      there is not enough evidence to decide between $H_{0}$ and $H_{1}$.
\item[Lower Right:]
      Here $p_{1}$ is small and $B_{01}$ large.  Both support the rejection of
      $H_{1}$ in favor of $H_{0}$.
\end{description}
Curves of constant $\tau/\sigma$ represent fixed experimental conditions, such 
that repeated observations would fall randomly (but not necessarily uniformly)
along one such curve.  On a given curve there is agreement between $p$-values 
and Bayes factors at high and low $p_{0}$, but somewhere in between there is a 
region of either no-decision (low $\tau/\sigma$) or paradox (high $\tau/\sigma$).

%%%%%%%%%%%%%%%%%%%%%%%%%%%%%%%%%%%%%%%%%%%%%%%%%%%%%%%%%%%%%%%%%%%%%%%%%%%%%%%%
\subsubsection{Possible solutions to the paradox}
%%%%%%%%%%%%%%%%%%%%%%%%%%%%%%%%%%%%%%%%%%%%%%%%%%%%%%%%%%%%%%%%%%%%%%%%%%%%%%%%
Over the years many solutions have been proposed to the Jeffreys-Lindley 
paradox.  Here we briefly illustrate two arguments.

\begin{figure}[p]
\begin{center}
\includegraphics[width=\textwidth]{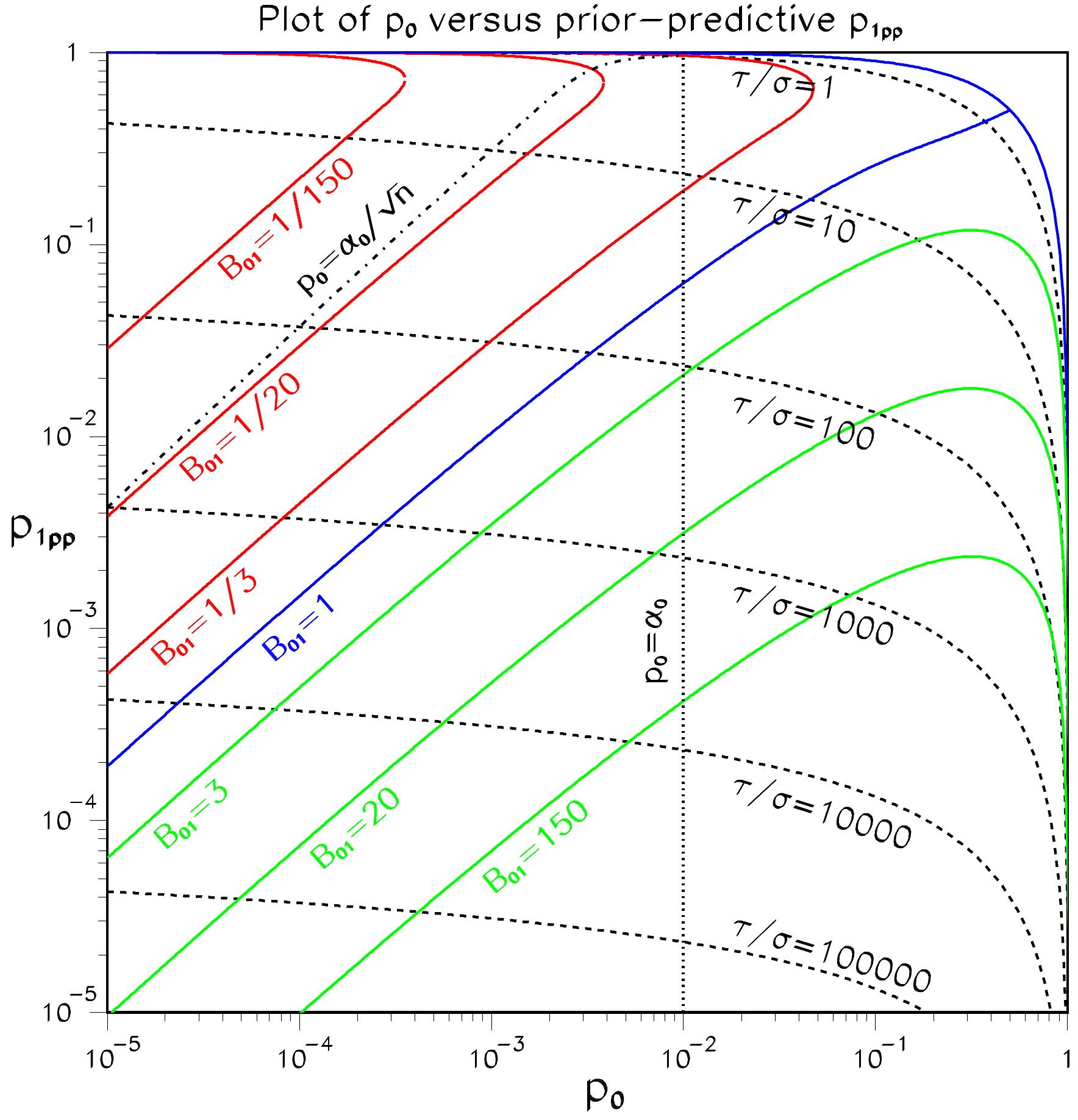}
\caption{
Log-log version of figure~\protect\ref{fig:ppp1_v_p0_lin}, with a larger range
of fixed-hypothesis contours and more realistic values for the Bayes factor
contours.  Also shown are the constant $p_{0}$ threshold (dotted line) at
$\alpha_{0}=1\%$ and the corresponding $n$-dependent threshold 
$\alpha_{0}/\sqrt{n}$ (dot-dashed line), where $n$ is the sample size.
\label{fig:ppp1_v_p0_log}}
\end{center}
\end{figure}

The first argument essentially blames the $p$-value method for the paradox and 
argues that with increasing values of $\tau/\sigma$ the $p$-value discovery 
threshold $\alpha_{0}$ should be lowered.  This argument is usually applied to 
the situation where $\sigma$ depends on a sample size $n$, so that $\tau/\sigma$
is proportional to $\sqrt{n}$.  In figure~\ref{fig:ppp1_v_p0_log} for example, 
one could think of the contours $\tau/\sigma=1$, $10$, $100$,\ldots as 
corresponding to $n=1$, $100$, $10\,000$,\dots, respectively.  If one chooses 
a discovery threshold of $1\%$ on the $\tau/\sigma=1$ contour, $0.1\%$ on the 
$\tau/\sigma=10$ contour, and so on, the dot-dashed curve labeled 
$p_{0}=\alpha_{0}/\sqrt{n}$ (where $\alpha_{0}$ is the discovery threshold on
the $\tau/\sigma=1$ contour) is obtained.  At large $\tau/\sigma$ this curve 
follows pretty closely the shape of the constant Bayes factor contours.  Thus, 
cutting on $p_{0}<\alpha_{0}/\sqrt{n}$ instead of $p_{0}<\alpha_{0}$ avoids the 
Jeffreys-Lindley paradox.  Interestingly, this is the same solution that was 
proposed to avoid sampling to a foregone conclusion in 
section~\ref{sec:foregone-conclusion}.

In a similar vein, it has been argued~\cite{Cox2014} that for experiments that 
collect more and more data, the realistic values of $\mu$ to be considered under
$H_{1}$ (assuming that no evidence for $\mu > \mu_{0}$ has been obtained, and we 
still believe that a small difference is possible) should be those that are 
closer and closer to $\mu_{0}$.  Thus the prior $\pi_{1}(\mu\mid\tau)$ in 
equation~\eqref{eq:H1-prior} should become narrower (smaller $\tau$), and this 
prevents $B_{01}$ favoring $H_{0}$ (as shown in figure~\ref{fig:ppp1_v_p0_log}).

\begin{figure}[p]
\begin{center}
\includegraphics[width=\textwidth]{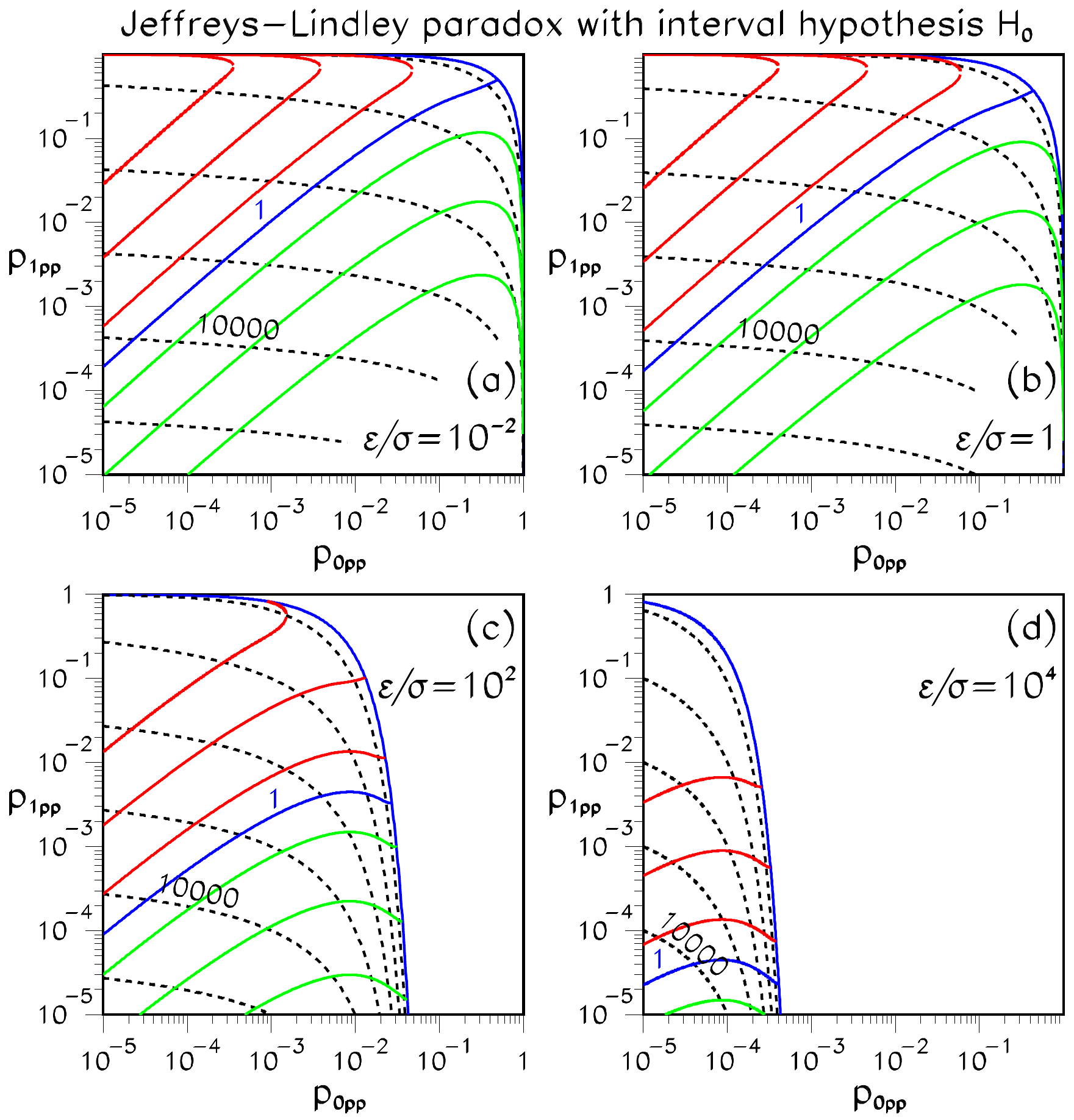}
\caption{
Plots illustrating what happens to the Jeffreys-Lindley paradox when $H_{0}$ is 
an interval hypothesis with width $\epsilon$ instead of a point-null hypothesis.  
Both $p_{0pp}$ and $p_{1pp}$ are prior-predictive $p$-values.  The contours on 
these plots are the same as in figure~\protect\ref{fig:ppp1_v_p0_log}, although 
for clarity only the contours $\tau/\sigma=10000$ and $B_{01}=1$ are labeled.
Compared with figure~\protect\ref{fig:ppp1_v_p0_log}, the fixed-hypothesis 
contours and the constant Bayes factor contours are both changed in such a way 
that the paradox remains present regardless of the value of $\epsilon/\sigma$. 
\label{fig:jlsolpp}}
\end{center}
\end{figure}

\begin{figure}[p]
\begin{center}
\includegraphics[width=\textwidth]{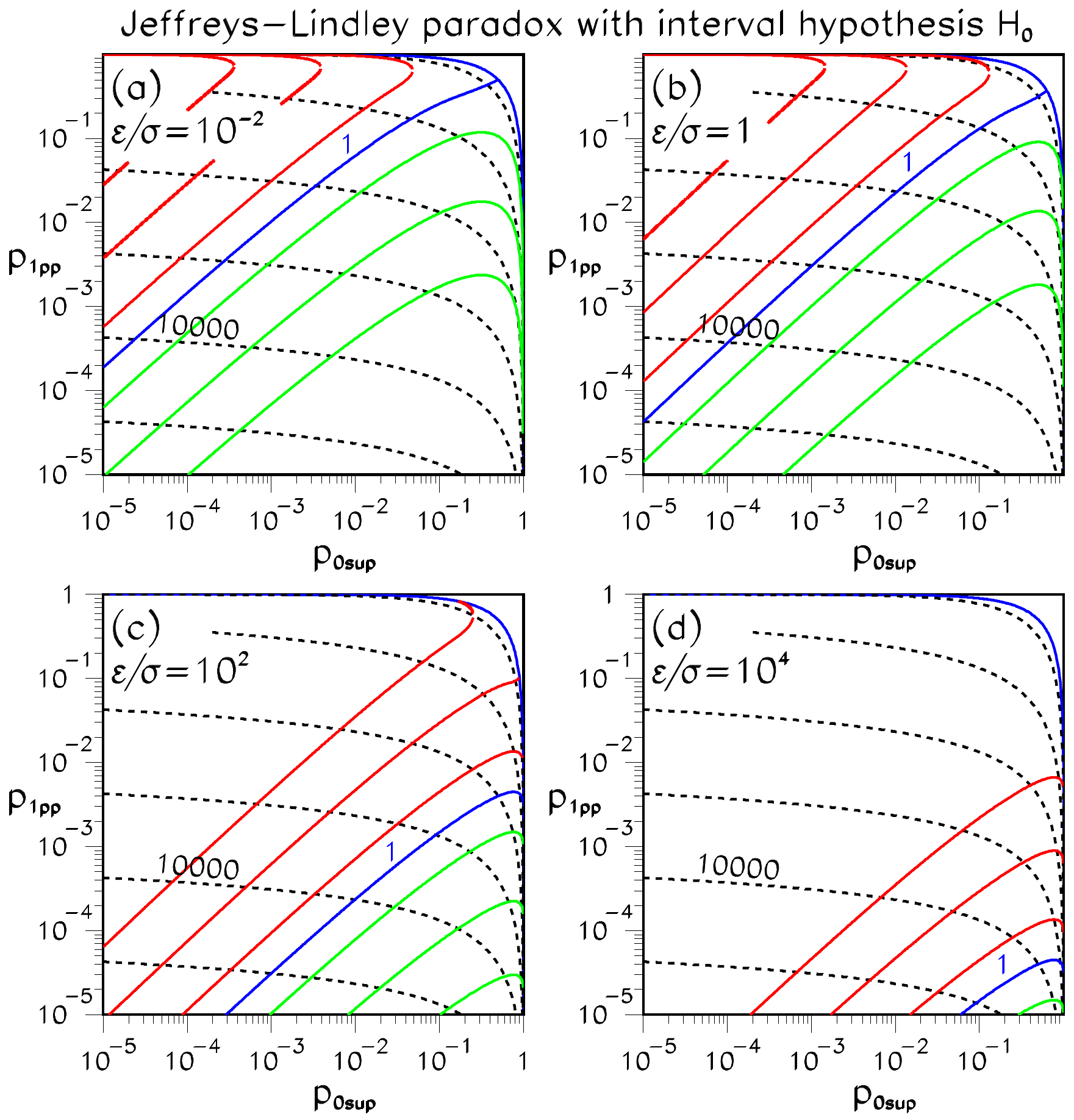}
\caption{
Plots illustrating what happens to the Jeffreys-Lindley paradox when $H_{0}$ is 
an interval hypothesis with width $\epsilon$ instead of a point-null hypothesis.  
The $p$-value $p_{1pp}$ is a prior-predictive $p$-value, whereas $p_{0\sup}$ is 
a supremum $p$-value.  The contours on these plots are the same as in 
figure~\protect\ref{fig:ppp1_v_p0_log}, although for clarity only the contours 
$\tau/\sigma=10000$ and $B_{01}=1$ are labeled.  Compared with 
figure~\protect\ref{fig:ppp1_v_p0_log}, only the constant Bayes factor contours 
are changed; the fixed-hypothesis contours are the same.  The result is that
the paradox disappears for a suitably high value of $\epsilon/\sigma$.
\label{fig:jlsolsup}}
\end{center}
\end{figure}

For the second argument, note that in the region of disagreement between $p_{0}$
and Bayes factors, both $p_{0}$ and $p_{1pp}$ tend to be small: one is in the 
double-rejection region of the test for most values of $\mu$ under $H_{1}$.  
This should alert the experimenter to the possibility that a third hypothesis 
may be true, or that there may be a mismodeling error.  One such error could be 
that $H_{0}$, rather than a point null hypothesis, is in fact an interval 
hypothesis with width $\epsilon$.  Thus, instead of~\eqref{eq:JL-testing-problem-1}, 
one should really be testing
\begin{equation}
H_{0}:\; \mu_{0}-\epsilon<\mu\le\mu_{0}\quad\textrm{ versus }\quad H_{1}:\; \mu>\mu_{0}.
\label{eq:JL-testing-problem-2}
\end{equation}
We consider two different regimes for $\epsilon$. The first has 
$\epsilon/\sigma=0.01$ or $1$, corresponding to a small or moderate widening 
of the original $H_{0}$.  The second regime uses $\epsilon/\sigma=100$ or 
$10^{4}$, which almost changes $H_{0}$ to $\mu\le\mu_{0}$, the complement of
$H_{1}:\mu>\mu_{0}$.  For both regimes one will need to introduce a prior 
$\pi_{0}(\mu)$ for $\mu$ under $H_{0}$, and the Bayes factor becomes:
\begin{equation}
B_{01}\;\equiv\;\frac{\int\! L_{0}(\mu)\,\pi_{0}(\mu)\,d\mu}
                     {\int\! L_{1}(\mu)\,\pi_{1}(\mu)\,d\mu}.
\label{eq:JL-Bayes-factor-2}
\end{equation}
For the $p$-value under $H_{0}$ one could again consider a prior-predictive
version:
\begin{equation}
p_{0pp}\;\equiv\; \int \! \pi_{0}(\mu) \int^{+\infty}_{t_{0}} f(t\mid\mu)\, dt\,d\mu,
\label{eq:prior-pred-p0}
\end{equation}
or choose a frequentist approach, such as the supremum $p$-value~\cite{Demortier2008}:
\begin{equation}
p_{0\sup}\;\equiv\; \sup_{\mu\in[\mu_{0}-\epsilon,\,\mu_{0}]}\; 
                \int^{+\infty}_{t_{0}} f(t\mid\mu)\, dt.
\label{eq:supremum-p0}
\end{equation}
Figures~\ref{fig:jlsolpp} and~\ref{fig:jlsolsup} illustrate the effect of these 
two definitions on the Jeffreys-Lindley paradox.  Note first that in both cases
one recovers figure~\ref{fig:ppp1_v_p0_log} when $\epsilon/\sigma$ is small.
When the prior-predictive $p_{0pp}$ of equation~\eqref{eq:prior-pred-p0} is used
(figure~\ref{fig:jlsolpp}), a given observation above $\mu_{0}$ becomes more 
significant since its $p_{0}$-value is averaged over $\mu$ values below 
$\mu_{0}$.  This causes the fixed-hypothesis contours to be compressed towards 
low $p_{0}$.  At the same time, the Bayes factor of such an observation
tends to decrease due to the numerator being replaced by an average; this 
causes the constant Bayes factor contours to move down.  The net effect of 
these contour changes is to leave the paradox in place.  This can be seen, for
example, by considering the point with $p_{0}=10^{-4}$ on the $\tau/\sigma=10\,000$
contour.  In all four plots of figure~\ref{fig:jlsolpp} this point hardly moves, 
having a Bayes factor $B_{01}$ close to 3.

On the other hand, when the supremum $p$-value of equation~\eqref{eq:supremum-p0}
is used for $p_{0}$ (figure~\ref{fig:jlsolsup}), only the constant Bayes factor 
contours change.  The fixed-hypothesis contours stay the same, because the 
supremum of $p_{0}$ over the interval $[\mu_{0}-\epsilon,\,\mu_{0}]$ is attained 
at $\mu=\mu_{0}$\footnote{Note that the supremum of $p_{1}$ over the interval 
$]\mu_{0},\mu_{0}+\tau]$ is also attained at $\mu=\mu_{0}$, so that 
$p_{0\sup}+p_{1\sup}=1$ (for continuous pdf's).  Therefore there is nothing to 
be learned from a plot of $p_{0\sup}$ versus $p_{1\sup}$.}.  For 
fixed $\tau/\sigma$, increasing $\epsilon/\sigma$ causes the paradoxical region 
to be pushed toward larger values of $p_{0}$ and smaller values of $p_{1}$.  
Eventually the $p$-values agree with the Bayes factor and the paradox disappears.  
In principle one could even tune the value of $\epsilon/\sigma$ to obtain 
$B_{01}=1$ at a specified value of $p_{0}$ (keeping $\tau/\sigma$ constant).  
Smaller values of $p_{0}$ would then correspond to $B_{01}$ disfavoring $H_{0}$, 
and larger $p_{0}$ to $B_{01}$ favoring $H_{0}$.

We conclude from this discussion of the second argument that introduction of a 
scale $\epsilon$ under $H_{0}$ is by itself not sufficient to suppress the 
paradox.  One also needs to specify how to handle $\epsilon$ in the computation 
of $p_{0}$.  Furthermore, as shown in figure~\ref{fig:jlsolsup} the paradox is 
not fully suppressed unless $\epsilon/\sigma$ is substantially larger than 1,
of the same order as $\tau/\sigma$.  Thus, the hierarchy $\epsilon\ll\sigma\ll\tau$,
presented in ref.~\cite{Cousins2013}, is sufficient to produce the paradox, but
not necessary.  When using the supremum $p$-value, the condition
[$\epsilon\ll\tau$ and $\sigma\ll\tau$] is both necessary and sufficient.

%%%%%%%%%%%%%%%%%%%%%%%%%%%%%%%%%%%%%%%%%%%%%%%%%%%%%%%%%%%%%%%%%%%%%%%%%%%%%%%%
\subsubsection{Simple versus simple version of the Jeffreys-Lindley paradox}
%%%%%%%%%%%%%%%%%%%%%%%%%%%%%%%%%%%%%%%%%%%%%%%%%%%%%%%%%%%%%%%%%%%%%%%%%%%%%%%%
Figures~\ref{fig:ppp1_v_p0_lin} and~\ref{fig:ppp1_v_p0_log} do not look very 
different from figures~\ref{fig:likelihood_ratio} and~\ref{fig:likelihood_ratio_loglog}
discussed in the sections on likelihood ratios, in spite of the use of a 
different $p_{1}$ definition.  This is a consequence of the fact that the 
Jeffreys-Lindley paradox can be reformulated in the context of a simple versus 
simple test~\footnote{The simple versus simple scenario outlined in this section
is unrelated to the basic insight described in 
section~\protect\ref{sec:JL-Basic-Insight}.}.  As noted at the beginning of 
section~\ref{sec:JL-paradox}, the paradox occurs for tests of the form:
\begin{equation}
\textrm{Test 1:}\quad H_{0}: \mu=\mu_{0} \;\textrm{versus}\; H_{1}: \mu>\mu_{0},
\end{equation}
using a test statistic $T\sim f(t\mid\mu)$ and assuming a prior 
$\pi_{1}(\mu\mid\tau)$ for $\mu$ under $H_{1}$, where $\tau$ characterizes
the scale of $\pi_{1}$.%%%

To proceed with the reformulation, introduce a variate $X$ whose randomness is 
the result of a two-step generating process: $X\sim f(x\mid\mu)$, where 
$\mu\sim \pi_{1}(\mu\mid\theta)$.  Thus, for fixed $\theta$ the distribution of 
$X$ is:
\begin{equation}
\tilde{f}(x\mid\theta)\;=\; \int\!\pi_{1}(\mu\mid\theta)\,f(x\mid\mu)\,d\mu.
\end{equation}
If for example $f(x\mid\mu)$ is Gaussian with mean $\mu$ and standard
deviation $\sigma$, and $\pi_{1}(\mu\mid\theta)$ is the indicator function of
the interval $[\mu_{0},\mu_{0}+\theta]$, this will yield:
\begin{equation}
\tilde{f}(x\mid\theta)\;=\;
\frac{1}{2\,\theta}\left[\textrm{erf}\left(\frac{\mu_{0}+\theta-x}{\sqrt{2}\,\sigma}\right)
-\textrm{erf}\left(\frac{\mu_{0}-x}{\sqrt{2}\,\sigma}\right)\right].
\label{eq:JL-integrated-pdf}
\end{equation}
As $\theta\rightarrow 0$ this pdf approaches $f(x\mid\mu_{0})$, and we will 
assume that this remains true for any choice of prior $\pi_{1}$ (i.e., that for 
$\theta=0$, $\pi_{1}(\mu\mid\theta)$ is a delta function at $\mu=\mu_{0}$).  

\begin{figure}[p]
\begin{center}
\includegraphics[width=\textwidth]{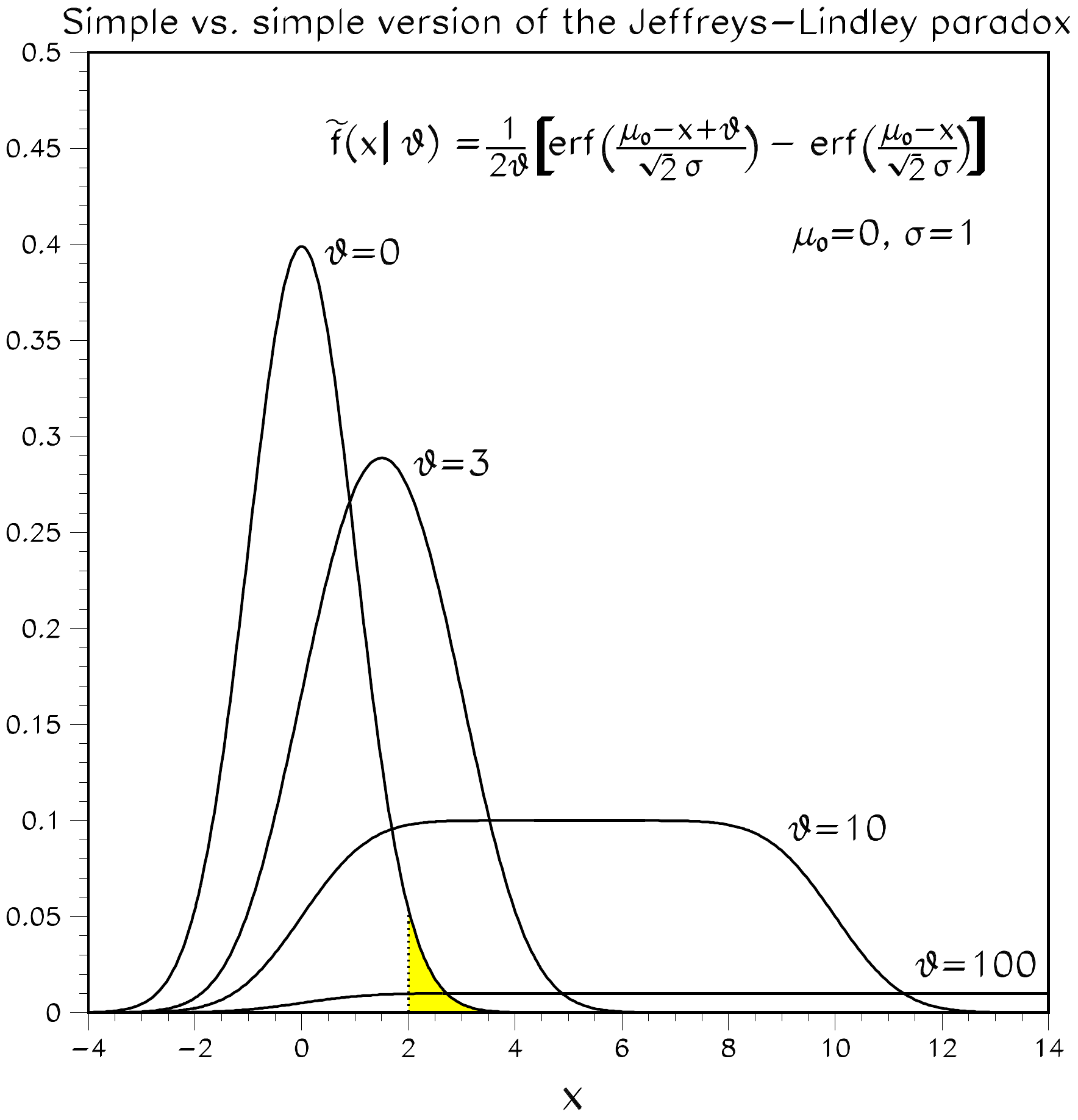}
\caption{
Plot of the integrated pdf~\protect\eqref{eq:JL-integrated-pdf} for several values
of the parameter $\theta$ (the pdf for $\theta=100$ has been truncated at the upper
end).  If for example $x=2$ is observed, the $p$-value under $H_{0}: \theta=0$ is 
2.3\% (shaded area), but the likelihood ratio of $\theta=0$ to $\theta=100$ is
5.5.  It is clear that for very large $\theta$ values, significantly small $p$-values
that disfavor $H_{0}$ will be associated with likelihood ratios that favor $H_{0}$.
This is a simple versus simple version of the Jeffreys-Lindley paradox.
\label{fig:taupdf}}
\end{center}
\end{figure}

Consider now the simple versus simple test:
\begin{equation}
\textrm{Test 2:}\quad H_{0}: \theta=0 \;\textrm{versus}\; H_{1}: \theta=\tau,
\end{equation}
using the test statistic $X\sim \tilde{f}(x\mid\theta)$.  Test~2 is designed to
determine whether or not the additional source of randomization $\pi_{1}$ is 
present in the process that generates $X$.  If $\theta=0$, there is no 
additional randomization and $\mu=\mu_{0}$.  On the other hand, if $\theta=\tau$,
additional randomization is present, its magnitude agrees with the prediction 
under $H_{1}$ in Test~1, and we must have $\mu>\mu_{0}$.  Tests~1 and~2 yield
the same information about $\mu$.  However, since $H_{1}$ is composite in Test~1 
but simple in Test~2, this has some interesting consequences.  The $p$-value 
$p_{1}$ is prior-predictive in Test~1 but standard frequentist in Test~2
(as can be seen by interchanging the order of integration in 
equation~\eqref{eq:prior-pred-p1}).  The Bayes factor in Test~1 is a likelihood
ratio for Test~2.  Tests~1 and~2 yield the same $p_{0}$ versus $p_{1}$ plots.
The Jeffreys-Lindley paradox, which is a disagreement between $p$-values and
Bayes factors in Test~1, is a disagreement between $p$-values and likelihood
ratios in Test~2.  This purely frequentist version of the Jeffreys-Lindley 
paradox is illustrated in figure~\ref{fig:taupdf} using the pdf 
$\tilde{f}(x\mid\theta)$ of equation~\eqref{eq:JL-integrated-pdf}.  It
shows that when testing a narrow distribution against a very broad one, it is 
possible to observe data with small $p$-value under the narrow-distribution 
hypothesis and yet large likelihood ratio in favor of that hypothesis.

Even though Test~1 is not of the simple versus simple type, it is possible to
define a likelihood ratio statistic for it, as the ratio of the likelihood
under $H_{0}$ to the maximized likelihood under $H_{1}$, where the maximum is
taken over $\mu > \mu_{0}$.  An interesting quantity is the Ockham factor, 
defined as the ratio of the Bayes factor to this likelihood ratio.  For Test~1
the Ockham factor is approximately $\tau/(\sqrt{2\pi}\,\sigma)$.  This is 
approximately proportional to the ratio of the widths of the distributions under 
$H_{1}$ and $H_{0}$ in Test~2.  More interestingly, at large $\tau/\sigma$ the 
Type-II error rate of the simple versus simple test equals $N_{\sigma}\sigma/\tau$,
where $N_{\sigma}$ is the number of standard deviations corresponding to the
cutoff $\alpha_{0}$ used to reject $H_{0}$.  Hence the Type-II error rate is
inversely proportional to the Ockham factor: if the alternative hypothesis is
true, the probability of rejecting the null with $p_{0}$ increases with $\tau/\sigma$,
but so does the disagreement between $p_{0}$ and $B_{01}$!  Referring again to 
figure~\ref{fig:taupdf}, we see that for small $x$ values (say below $x=2$), 
Bayes factors and $p$-values both favor $H_{0}$.  At high $x$ they both disfavor 
$H_{0}$.  In between there is a region where agreement between $p$-values and 
Bayes factors depends on the Ockham factor.

%%%%%%%%%%%%%%%%%%%%%%%%%%%%%%%%%%%%%%%%%%%%%%%%%%%%%%%%%%%%%%%%%%%%%%%%%%%%%%%%
\section{Nuisance parameters}
\label{sec:NuisanceParameters}
%%%%%%%%%%%%%%%%%%%%%%%%%%%%%%%%%%%%%%%%%%%%%%%%%%%%%%%%%%%%%%%%%%%%%%%%%%%%%%%%
There are many methods for eliminating nuisance parameters from $p$-value 
calculations (see for example~\cite{Demortier2008}), and the choice of method 
will generally have an effect on the construction of $p_{0}$ versus $p_{1}$ 
plots.  We start with a couple of examples.

First consider the situation where one makes $n$ measurements $x_{i}$ from a 
Gaussian population with unknown mean $\mu$ and unknown width $\sigma$.  
A sufficient statistic consists of the pair $(\bar{x},s)$, where 
$\bar{x} = \frac{1}{n}\sum_{i=1}^{n}x_{i}$ is the sample mean and 
$s = [\frac{1}{n-1}\sum_{i=1}^{n}(x_{i}-\bar{x})^{2}]^{1/2}$ is the sample
standard deviation.  To test the hypotheses $H_{i}: \mu=\mu_{i}$ ($i=0,1$), the
classical approach uses the test statistics $t_{i} = \sqrt{n}\,(\bar{x}-\mu_{i})/s$, 
which have Student's $t$ distribution under the respective $H_{i}$.  Thus one can 
calculate the $p$-values $p_{0}$ and $p_{1}$.  Unfortunately the relation 
between $p_{0}$ and $p_{1}$ is not one-to-one: from $p_{0}$ one can obtain 
$t_{0}$, but from $t_{0}$ one cannot extract both $\bar{x}$ and $s$, which are
needed to compute $t_{1}$ and then $p_{1}$.  Hence it is not possible to make
a plot of $p_{0}$ versus $p_{1}$.  This problem is related to the fact 
that the power of the $t$ test depends on the unknown value of $\sigma$ and 
not just on the significance threshold $\alpha$.

For the second example we consider the observation of a Poisson variate $N$,
whose mean is the product of a parameter of interest $\mu$ and a nuisance
parameter $\kappa$.  Again we wish to test $H_{0}: \mu=\mu_{0}$ versus
$H_{1}: \mu=\mu_{1}$.  Information about the nuisance parameter comes from a 
second Poisson measurement $K$, with mean $\kappa$.  A well-known approach for 
this case is to condition on the sum $N+K$.  The distribution of $N$, given a 
fixed value of $N+K$, is binomial with parameter $\mu/(1+\mu)$.  Here one can 
calculate conditional $p$-values $p_{0}$ and $p_{1}$, and plot one against the 
other.

The above examples rely on a special structure of the problem under study to
eliminate nuisance parameters.  Unfortunately such a special structure is not always
available, and even when it is, it does not guarantee that a $(p_{0},p_{1})$ 
plot can be constructed.  Here we offer a couple of suggestions for handling
the general case.  The first one is to use parametric bootstrap techniques to 
eliminate the nuisance parameters.  To first order these techniques consist in 
substituting an estimate for the unknown nuisance parameter values.  The 
resulting $p$-values are generally no longer uniform under their respective null 
hypothesis, but there exist higher-order refinements that restore some of that 
uniformity~\cite{bootstrap}.  Bootstrap computations can quickly become 
rather intensive, but they have the advantage of being frequentist and therefore 
preserving the error structure of the tests discussed in section~\ref{sec:ErrorRates}.
As for the likelihood ratios, they can be replaced by profile likelihood ratios.
Although the latter are not genuine likelihood ratios, with some caveats they 
can still be treated as representing statistical evidence in large 
samples~\cite{Royall2000}.

Our second suggestion is to apply Bayesian methods on the nuisance 
parameters.  Effectively, this amounts to replacing composite hypotheses by
simple ones, by integrating out the nuisance parameters over an appropriate
proper prior.  Suppose for example that the probability density of the data 
$x$ under $H_{i}$ is given by $f(x\mid \mu_{i},\nu)$, with $\nu$ a vector of 
nuisance parameters with prior $\pi(\nu)$.  Then we simply replace $f$ by
\begin{equation}
f^{\star}(x\mid\mu_{i})\;=\;\int f(x\mid\mu_{i},\nu)\,\pi(\nu)\,d\nu
\end{equation}
in the formulation of the hypotheses.  The $p$-values become {\em prior-predictive}
$p$-values:
\begin{multline}
p_{i}^{\star}\;=\; \int_{x_{0}}^{\infty}f^{\star}(x\mid\mu_{i})\,dx
             \;=\; \int_{x_{0}}^{\infty}\int f(x\mid\mu_{i},\nu)\,\pi(\nu)\,d\nu\,dx\\[2mm]
             \;=\; \int \left[\int_{x_{0}}^{\infty} f(x\mid\mu_{i},\nu)\,dx\right]\pi(\nu)\,d\nu
             \;=\; \int p_{i}(\nu)\,\pi(\nu)\,d\nu
\end{multline}
(compare equation~\eqref{eq:prior-pred-p1}), and the likelihood ratios become 
Bayes factors:
\begin{equation}
\lambda_{01}\;=\;\frac{f^{\star}(x_{0}\mid\mu_{0})}{f^{\star}(x_{0}\mid\mu_{1})}
            \;=\;\frac{\int f(x_{0}\mid\mu_{0},\nu)\,\pi(\nu)\,d\nu}
                      {\int f(x_{0}\mid\mu_{1},\nu)\,\pi(\nu)\,d\nu}.
\end{equation}
Although this approach lacks the frequentist error interpretation of the tests,
it still enjoys the evidential interpretation of the $p$-values and Bayes 
factors.  It is also conceptually simpler and more elegant, as well as 
computationally much easier, than the bootstrap.

%%%%%%%%%%%%%%%%%%%%%%%%%%%%%%%%%%%%%%%%%%%%%%%%%%%%%%%%%%%%%%%%%%%%%%%%%%%%%%%%
\section{Conclusion}
\label{sec:Conclusion}
%%%%%%%%%%%%%%%%%%%%%%%%%%%%%%%%%%%%%%%%%%%%%%%%%%%%%%%%%%%%%%%%%%%%%%%%%%%%%%%%
We find that $(p_{0},p_{1})$ plots such as figs.~\ref{fig:contours}, 
\ref{fig:likelihood_ratio} and~\ref{fig:likelihood_ratio_loglog} provide useful 
insights into several diverse statistical aspects of searches for new physics:
\begin{itemize}
\item The $CL_{s}$ criterion for exclusing $H_{1}$;
\item The Punzi definition of sensitivity;
\item The relationship between $p$-values and likelihoods;
\item The difference between the exclusion regions using $p$-values and likelihoods;
\item The probabilities of misleading evidence;
\item The Jeffreys-Lindley paradox.
\end{itemize}
In addition, we believe that these plots could be helpful in summarizing the 
results of such searches.  When these involve many channels, with possibly 
different sensitivities, one could plot the results as points on a 
$(p_{0},p_{1})$ plot, together with Gaussian likelihood-ratio contours (since
the latter are large-sample limits of the actual data pdf's).  This would 
provide a convenient graphical overview of both the $p$-value and the 
likelihood-ratio evidence contained in the ensemble of channels investigated.

%%%%%%%%%%%%%%%%%%%%%%%%%%%%%%%%%%%%%%%%%%%%%%%%%%%%%%%%%%%%%%%%%%%%%%%%%%%%%%%%
\section{Acknowledgments}
%%%%%%%%%%%%%%%%%%%%%%%%%%%%%%%%%%%%%%%%%%%%%%%%%%%%%%%%%%%%%%%%%%%%%%%%%%%%%%%%
We are very grateful to Sir David Cox for insights into the Jeffreys-Lindley 
paradox, and to Bob Cousins for interesting dicussions about it.

%%%%%%%%%%%%%%%%%%%%%%%%%%%%%%%%%%%%%%%%%%%%%%%%%%%%%%%%%%%%%%%%%%%%%%%%%%%%%%%%

%%%%%%%%%%%%%%%%%%%%%%%%%%%%%%%%%%%%%%%%%%%%%%%%%%%%%%%%%%%%%%%%%%%%%%%%%%%%%%%%
\newpage
\appendix
\section{The \texorpdfstring{Bayes-$\mathbf{CL_{s}}$}{Bayes-CLs} connection}
\label{AppA}
This appendix describes a sufficient condition for $CL_{s}$ upper limits to 
agree with Bayesian upper limits.

Let $f(x\mid \mu)$ be a family of probability densities for the random variable
$X$, indexed by the parameter $\mu$, and consider the family of tests:
\begin{equation}
H_{0}[\mu^{\star}]:\;\mu=\mu^{\star} \quad\textrm{ versus }\quad 
H_{1}[\mu^{\star}]:\;\mu>\mu^{\star}.
\end{equation}
Suppose we observe $X=x_{0}$.  If we have a prior $\pi(\mu)$ for $\mu$ under 
$H_{1}[\mu^{\star}]$, the Bayesian evidence in favor of $H_{1}[\mu^{\star}]$ is 
simply the marginal probability of $x_{0}$ under $H_{1}[\mu^{\star}]$:
\begin{equation}
p(x_{0}\mid H_{1}[\mu^{\star}]) \;=\; \int_{\mu^{\star}}^{+\infty} 
                                      f(x_{0}\mid\mu)\,\pi(\mu)\,d\mu.
\end{equation}
Note that this probability is only correctly normalized if $\pi(\mu)$ is a 
proper prior.  However, the argument that follows remains valid if $\pi(\mu)$
is improper.  The $p$-value evidence against $H_{0}[\mu^{\star}]$, when the 
alternative is $H_{1}[\mu^{\star}]$, is:
\begin{equation}
p_{0}(\mu^{\star}) \;=\; \int_{x_{0}}^{+\infty} f(x\mid\mu^{\star})\, dx 
                   \;=\; 1 - F(x_{0}\mid\mu^{\star}),
\end{equation}
where $F(x\mid\mu)$ is the cumulative probability distribution of $x$.  This
$p$-value evidence against $H_{0}[\mu^{\star}]$ increases as 
$1-p_{0}(\mu^{\star})$ increases.  Assume now that the Bayesian and frequentist 
evidences are equal for all $\mu^{\star}$ values larger than some prespecified 
$\mu_{0}$:
\begin{equation}
p(x_{0}\mid H_{1}[\mu^{\star}]) \;=\; 1-p_{0}(\mu^{\star}), 
\textrm{ for all $\mu^{\star}\ge\mu_{0}$},
\label{eq:XIntMuInt}
\end{equation}
or:
\begin{equation}
\int_{\mu^{\star}}^{+\infty} f(x_{0}\mid\mu)\,\pi(\mu)\, d\mu \;=\;
\int_{-\infty}^{x_{0}} f(x\mid\mu^{\star})\,dx, 
\textrm{ for all $\mu^{\star}\ge\mu_{0}$}.
\end{equation}
This condition is sufficient to obtain equality of $CL_{s}$ and Bayesian upper
limits on $\mu$ under $H_{1}[\mu_{0}]$.  Indeed, the $\gamma$-credibility level 
upper limit $\mu_{U}$ on $\mu$ is the solution of:
\begin{equation}
\int_{\mu_{0}}^{\mu_{U}}\!p(\mu\mid x_{0},H_{1}[\mu_{0}])\;d\mu\;=\;\gamma,
\label{eq:Ulim0}
\end{equation}
where the integrand is the posterior density of $\mu$ under $H_{1}[\mu_{0}]$:
\begin{equation}
p(\mu\mid x_{0},H_{1}[\mu_{0}])
\;=\;\frac{f(x_{0}\mid\mu)\;\pi(\mu)}
{\int_{\mu_{0}}^{+\infty}\!f(x_{0}\mid\mu^{\prime})\;\pi(\mu^{\prime})\;d\mu^{\prime}}
\;=\;\frac{f(x_{0}\mid\mu)\;\pi(\mu)}{p(x_{0}\mid H_{1}[\mu_{0}])}.
\label{eq:MuPosterior}
\end{equation}
Substituting equation~\eqref{eq:MuPosterior} in~\eqref{eq:Ulim0} leads to:
\begin{equation}
\frac{p(x_{0}\mid H_{1}[\mu_{0}]) - p(x_{0}\mid H_{1}[\mu_{U}])}
     {p(x_{0}\mid H_{1}[\mu_{0}])} \; =\; \gamma,
\end{equation}
and using condition~\eqref{eq:XIntMuInt} yields:
\begin{equation}
1 - \frac{1-p_{0}(\mu_{U})}{1-p_{0}(\mu_{0})}\;=\; \gamma.
\label{eq:Ulim1}
\end{equation}
The quantity $1-p_{0}(\mu_{U})$ is in fact $p_{1}(\mu_{U})$, the $p$-value for
testing $\mu=\mu_{U}$ when the alternative is $\mu=\mu_{0}$.  Hence,
equation~\eqref{eq:Ulim1} is equivalent to
\begin{equation}
\frac{p_{1}(\mu_{U})}{1-p_{0}(\mu_{0})}\;=\;1-\gamma,
\label{eq:Ulim2}
\end{equation}
which corresponds to the $CL_{s}$ construction of upper limits.

We illustrate this result with two examples of families of distributions that 
satisfy condition~\eqref{eq:XIntMuInt}.  The first one is any family of 
continuous distributions parametrized by a location parameter:
\begin{equation}
f(x\mid\mu)\;=\; f(x\,-\,\mu).
\end{equation}
It is straightforward to verify, by integration by substitution, that
\begin{equation}
\int_{-\infty}^{x}\!f(x^{\prime}\,-\,\mu)\;dx^{\prime}
\;=\;\int_{\mu}^{+\infty}\!f(x\,-\,\mu^{\prime})\;d\mu^{\prime},
\end{equation}
so that condition~\eqref{eq:XIntMuInt} is indeed satisfied for a flat prior, 
$\pi(\mu)=1$.  The second example is the Poisson family:
\begin{equation}
f(n\mid\mu)\;=\;\frac{\mu^{n}}{n!}\,e^{-\mu},
\end{equation}
for which we have:
\begin{equation}
\sum_{i=0}^{n}\frac{\mu^{i}}{i!}\,e^{-\mu}\;=\;\int_{\mu}^{+\infty}\frac{t^{n}}{n!}\,e^{-t}\;dt,
\end{equation}
as can be checked by repeated integration by parts of the right-hand side.
Although this is a discrete version of condition~\eqref{eq:XIntMuInt}, also
with a flat prior, nothing essential changes in the argument leading 
from~\eqref{eq:Ulim0} to~\eqref{eq:Ulim2}.
\vfill
%%%%%%%%%%%%%%%%%%%%%%%%%%%%%%%%%%%%%%%%%%%%%%%%%%%%%%%%%%%%%%%%%%%%%%%%%%%%%%%%
\section{Figures}
%%%%%%%%%%%%%%%%%%%%%%%%%%%%%%%%%%%%%%%%%%%%%%%%%%%%%%%%%%%%%%%%%%%%%%%%%%%%%%%%
%%%%%%%%%%%%%%%%%%%%%%%%%%%%%%%%%%%%%%%%%%%%%%%%%%%%%%%%%%%%%%%%%%%%%%%%%%%%%%%%
%%%%%%%%%%%%%%%%%%%%%%%%%%%%%%%%%%%%%%%%%%%%%%%%%%%%%%%%%%%%%%%%%%%%%%%%%%%%%%%%
%%%%%%%%%%%%%%%%%%%%%%%%%%%%%%%%%%%%%%%%%%%%%%%%%%%%%%%%%%%%%%%%%%%%%%%%%%%%%%%%
%%%%%%%%%%%%%%%%%%%%%%%%%%%%%%%%%%%%%%%%%%%%%%%%%%%%%%%%%%%%%%%%%%%%%%%%%%%%%%%%
%%%%%%%%%%%%%%%%%%%%%%%%%%%%%%%%%%%%%%%%%%%%%%%%%%%%%%%%%%%%%%%%%%%%%%%%%%%%%%%%
%%%%%%%%%%%%%%%%%%%%%%%%%%%%%%%%%%%%%%%%%%%%%%%%%%%%%%%%%%%%%%%%%%%%%%%%%%%%%%%%
%%%%%%%%%%%%%%%%%%%%%%%%%%%%%%%%%%%%%%%%%%%%%%%%%%%%%%%%%%%%%%%%%%%%%%%%%%%%%%%%
%%%%%%%%%%%%%%%%%%%%%%%%%%%%%%%%%%%%%%%%%%%%%%%%%%%%%%%%%%%%%%%%%%%%%%%%%%%%%%%%
%%%%%%%%%%%%%%%%%%%%%%%%%%%%%%%%%%%%%%%%%%%%%%%%%%%%%%%%%%%%%%%%%%%%%%%%%%%%%%%%
\vspace*{-10mm}
\begin{table}[H]
\begin{center}
\begin{tabular}{|lcc|}
\hline
$\Delta\mu/\sigma$                          & 1.67   & 3.33   \\
\hline\hline
$\mathbb{P}(L_{0}/L_{1} < 1/32 \mid H_{0})$ & 0.18\% & 0.34\% \\
$\mathbb{P}(L_{0}/L_{1} < 1/8  \mid H_{0})$ & 1.9\%  & 1.1\%  \\
$\mathbb{P}(L_{0}/L_{1} >  8   \mid H_{1})$ & 1.9\%  & 1.1\%  \\
$\mathbb{P}(L_{0}/L_{1} > 32   \mid H_{1})$ & 0.18\% & 0.34\% \\
\hline
\end{tabular}
\end{center}
\end{table}
%%%%%%%%%%%%%%%%%%%%%%%%%%%%%%%%%%%%%%%%%%%%%%%%%%%%%%%%%%%%%%%%%%%%%%%%%%%%%%%%
%%%%%%%%%%%%%%%%%%%%%%%%%%%%%%%%%%%%%%%%%%%%%%%%%%%%%%%%%%%%%%%%%%%%%%%%%%%%%%%%
%%%%%%%%%%%%%%%%%%%%%%%%%%%%%%%%%%%%%%%%%%%%%%%%%%%%%%%%%%%%%%%%%%%%%%%%%%%%%%%%
%%%%%%%%%%%%%%%%%%%%%%%%%%%%%%%%%%%%%%%%%%%%%%%%%%%%%%%%%%%%%%%%%%%%%%%%%%%%%%%%
%%%%%%%%%%%%%%%%%%%%%%%%%%%%%%%%%%%%%%%%%%%%%%%%%%%%%%%%%%%%%%%%%%%%%%%%%%%%%%%%
%%%%%%%%%%%%%%%%%%%%%%%%%%%%%%%%%%%%%%%%%%%%%%%%%%%%%%%%%%%%%%%%%%%%%%%%%%%%%%%%
%%%%%%%%%%%%%%%%%%%%%%%%%%%%%%%%%%%%%%%%%%%%%%%%%%%%%%%%%%%%%%%%%%%%%%%%%%%%%%%%
%%%%%%%%%%%%%%%%%%%%%%%%%%%%%%%%%%%%%%%%%%%%%%%%%%%%%%%%%%%%%%%%%%%%%%%%%%%%%%%%
%%%%%%%%%%%%%%%%%%%%%%%%%%%%%%%%%%%%%%%%%%%%%%%%%%%%%%%%%%%%%%%%%%%%%%%%%%%%%%%%
\end{document}